\documentclass[english,floatfix,twocolumn,showpacs, aps, prx, reprint, superscriptaddress, longbibliography]{revtex4-2}
\usepackage{amsmath}
\usepackage{hyperref}
\hypersetup{colorlinks=true, citecolor=blue, urlcolor=blue, linkcolor=blue, breaklinks=true}
\usepackage{amssymb}
\usepackage{graphicx, xcolor}
\newcommand{\commentout}[1]{}
\usepackage{booktabs, footnote}
\usepackage{physics}

\usepackage{orcidlink}

\begin{document}
\title{Analog quantum simulation of Shastry-Sutherland Ising materials}
\author{Vahagn Mkhitaryan$^\dagger$}
\thanks{These authors contributed equally.\\
$^\dagger$Email address:
\href{mailto:vmkhitar@purdue.edu}{vmkhitar@purdue.edu};
\href{mailto:mkornjaca@quera.com}{mkornjaca@quera.com};
\href{mailto:arnabb@purdue.edu}{arnabb@purdue.edu};
\href{mailto:shalaev@purdue.edu}{shalaev@purdue.edu};
\href{mailto:swang@quera.com}{swang@quera.com}}
\affiliation{Elmore Family School of Electrical and Computer Engineering, Purdue University, West Lafayette, IN 47909, USA}
\author{Milan Kornjača$^\dagger$}
\thanks{These authors contributed equally.\\
$^\dagger$Email address:
\href{mailto:vmkhitar@purdue.edu}{vmkhitar@purdue.edu};
\href{mailto:mkornjaca@quera.com}{mkornjaca@quera.com};
\href{mailto:arnabb@purdue.edu}{arnabb@purdue.edu};
\href{mailto:shalaev@purdue.edu}{shalaev@purdue.edu};
\href{mailto:swang@quera.com}{swang@quera.com}}
\affiliation{QuEra Computing Inc., 1380 Soldiers Field Road, Boston, MA, 02135, USA}
\author{Mao Lin}
\affiliation{Amazon Braket, Seattle, WA, 98170, United States}
\author{Alexander Schuckert}
\affiliation{QuEra Computing Inc., 1380 Soldiers Field Road, Boston, MA, 02135, USA}
\affiliation{QuEra Computing UK Ltd., Unit 2, Tech Foundry 2, Didcot OX11 0QQ, UK}
\affiliation{DIENS, École Normale Supérieure, PSL University, CNRS, INRIA, 45 rue d’Ulm, Paris 75005, France}
\author{Majd Hamdan}
\author{Sergio H. Cant\'u}
\affiliation{QuEra Computing Inc., 1380 Soldiers Field Road, Boston, MA, 02135, USA}
\author{Ammar Ali}
\affiliation{Department of Physics and Astronomy, Purdue University, West Lafayette, IN 47906, USA}
\author{Tommaso Macr\'i}
\author{Nathan Gemelke}
\author{Alexander Keesling}
\affiliation{QuEra Computing Inc., 1380 Soldiers Field Road, Boston, MA, 02135, USA}
\author{Peter Komar}
\affiliation{Amazon Braket, Seattle, 98170, WA, United States}
\author{Travis~S.~Humble}
\affiliation{Quantum Science Center, Oak Ridge National Laboratory, 1 Bethel Valley Road, Oak Ridge, 37831, TN, USA}
\author{Alexei Bylinski}
\author{Pedro~L.~S.~Lopes}
\affiliation{QuEra Computing Inc., 1380 Soldiers Field Road, Boston, MA, 02135, USA}
\author{Fangli Liu}
\affiliation{QuEra Computing Inc., 1380 Soldiers Field Road, Boston, MA, 02135, USA}
\author{Alexandra Boltasseva}
\affiliation{Elmore Family School of Electrical and Computer Engineering, Purdue University, West Lafayette, IN 47909, USA}
\affiliation{Purdue Quantum Science and Engineering Institute, Purdue University, West Lafayette, IN 47907, USA}
\author{Mikhail D. Lukin}
\affiliation{Department of Physics, Harvard University, Cambridge, MA 02138, USA}
\author{Arnab Banerjee$^\dagger$}
\affiliation{Department of Physics and Astronomy, Purdue University, West Lafayette, IN 47906, USA}
\affiliation{Quantum Science Center, Oak Ridge National Laboratory, 1 Bethel Valley Road, Oak Ridge, 37831, TN, USA}
\affiliation{Purdue Quantum Science and Engineering Institute, Purdue University, West Lafayette, IN 47907, USA}
\author{Vladimir M. Shalaev$^\dagger$}
\affiliation{Elmore Family School of Electrical and Computer Engineering, Purdue University, West Lafayette, IN 47909, USA}
\affiliation{Department of Physics and Astronomy, Purdue University, West Lafayette, IN 47906, USA}
\affiliation{Purdue Quantum Science and Engineering Institute, Purdue University, West Lafayette, IN 47907, USA}
\author{Sheng-Tao Wang$^\dagger$}
\affiliation{QuEra Computing Inc., 1380 Soldiers Field Road, Boston, MA, 02135, USA}
\date{\today}

\begin{abstract}
Frustrated magnets host rich emergent behavior because competing interactions generate dense manifolds of low-energy states. The Ising Shastry-Sutherland model is a paradigmatic setting, motivated experimentally by Ising-like rare-earth tetraborides such as TmB$_4$ and ErB$_4$, whose field-induced magnetization plateaus have motivated extensive microscopic modeling. Yet how tunable quantum fluctuations reorganize this material-motivated phase diagram remains largely unexplored. Here, we realize a transverse- and longitudinal-field Ising model on the Shastry-Sutherland lattice using a programmable Rydberg-atom array. We map its quantum phase diagram using density matrix renormalization group calculations and probe corresponding signatures experimentally in arrays containing more than 110 atoms. Beyond the established classical phases, the calculations identify two regimes absent in the classical limit and stabilized at finite transverse field: a $2/5$ filling plateau whose finite stability region emerges from a classically degenerate phase boundary and a symmetry-preserving quantum-dimer regime replacing classically selected collinear order. Experiments on finite open arrays reveal signatures consistent with these regimes across three boundary-pinning conditions, with the resulting response maps agreeing qualitatively with the calculated phase diagram. Our results establish programmable Rydberg-atom arrays as tools for exploring frustrated, material-motivated Hamiltonians and motivate searches for new, quantum-driven magnetization plateaus and quantum-dimer behavior in ErB$_4$ under a transverse field or in strain-tuned TmB$_4$.

\end{abstract}

\maketitle

\section*{Introduction}

Competing interactions in frustrated magnets cannot generally be satisfied simultaneously, generating dense manifolds of low-energy states that are exceptionally sensitive to weak perturbations~\cite{Lacroix2011}. This sensitivity underlies phenomena ranging from fractional magnetization plateaus~\cite{Chubukov1991,Alicea2009} and fluctuation-selected order~\cite{Kawamura1985} to quantum spin liquids~\cite{Balents2010,Savary2017}. Ising models provide a particularly transparent setting because such complexity emerges even without continuous spin degrees of freedom~\cite{Wannier1950,Harris1997,Moessner2001}. A paradigmatic example is the Ising model on the Shastry-Sutherland lattice~\cite{Shastry1981}. For the minimal classical model, exact ground-state analysis shows that only the $1/3$ fractional magnetization plateau is stable over a finite field interval, while the phase boundaries host extensive degeneracies~\cite{Dublenych2012}. Longer-range interactions can lift these degeneracies and stabilize additional fractional plateaus~\cite{Dublenych2013,Dublenych2014}. Quantum Shastry-Sutherland systems, in turn, are known to host dimerized phases, fluctuation-induced plateaus, and unconventional critical behavior~\cite{Kageyama1999,Meng2008,Lee2019,Yang2022, Marshall2023, Liu2025,Yadav2025, Khundzakishvili2025}. The exactly characterized classical phase diagram therefore provides a precise baseline for asking how quantum fluctuations reorganize frustrated magnetism.

The rare-earth tetraborides TmB$_4$ and ErB$_4$ provide prominent material settings for Ising magnetism on the Shastry-Sutherland lattice~\cite{Schafer1976,Etourneau1985,Siemensmeyer2008,Matas2010,Flury2025}. In both compounds, the rare-earth ions form Shastry-Sutherland layers, while the crystal-field environment produces strongly Ising-like low-energy moments~\cite{Schafer1976,Etourneau1985,Ye2017,Flury2025}. TmB$_4$ exhibits N\'{e}el order at low field, ferrimagnetic order at higher field, and a sequence of narrow fractional magnetization plateaus—including $1/7$, $1/8$, and $1/9$—associated with stripe-like spin textures~\cite{Siemensmeyer2008,Trinh2018}. ErB$_4$ provides a complementary example, with collinear antiferromagnetic order at zero field and a field-induced $1/2$-magnetization plateau~\cite{Michimura2006,Flury2025}. These distinct plateau sequences expose both the power and the limitations of the minimal Ising model: Shastry-Sutherland frustration generates extensive degeneracies, while material-specific interactions select the phases realized in a particular compound~\cite{Farkas2010,Dublenych2012,Dublenych2013,Dublenych2014,Wierschem2013}. They also motivate the central question of this work: how do tunable quantum fluctuations reorganize this classically frustrated plateau landscape?

Isolating the effects of quantum fluctuations in these materials is difficult because the lattice geometry and microscopic couplings are fixed, while experimentally accessible controls cannot vary the relevant energy scales independently. Programmable quantum simulators offer a complementary approach in which geometry, interactions, and quantum dynamics can be controlled separately~\cite{Cirac2012,Georgescu2014}. Previous work has reproduced the classical phase structure of the Shastry-Sutherland Ising model on quantum-annealing hardware~\cite{Kairys2020}, and hardware-accelerated classical annealing~\cite{Jha2021}. Rydberg-atom arrays are particularly well suited to accessing the quantum regime: tweezer-defined atomic positions encode the lattice geometry, distance-dependent van der Waals interactions realize Ising couplings, and coherent laser driving provides a tunable transverse field~\cite{Labuhn2016,Bernien2017,aquila:2023}. These systems now enable quantitative studies of many-body phase structure and dynamics in two-dimensional arrays containing hundreds of atoms~\cite{Keesling2019,Samajdar2020, Samajdar2021,Bluvstein:2021,Ebadi2021,Scholl2021, Li2022,Chen2023, Kornjaca2023,Manovitz:2025,Hirsbrunner2025,Zhang2025,Darbha2025,Balewski2025,Mark2025}. Applied to the present problem, this combination makes it possible to realize the Shastry-Sutherland geometry directly and systematically vary the longitudinal field, transverse field, and ratio of competing interactions within a single experimental platform.

Here, we realize the transverse- and longitudinal-field Ising model on the Shastry-Sutherland lattice using a programmable neutral-atom array, with experiments performed on Aquila~\cite{aquila:2023} through Amazon Braket~\cite{AWSBraket2020}. The platform provides tunable longitudinal and transverse fields together with geometric control of the ratio of competing interactions, enabling controlled cuts through a three-dimensional parameter space. Using density matrix renormalization group calculations, we map the corresponding ground-state phase diagram. In addition to recovering the established classical phases, the calculations identify two regimes stabilized by quantum fluctuations: a finite-width $2/5$ filling plateau, together with its symmetry-related $3/5$ counterpart, near the boundary between the N\'{e}el and $1/3$($2/3$)-plateau phases, and a symmetry-preserving quantum-dimer regime replacing classically selected collinear order. We probe signatures of these regimes experimentally in finite open arrays containing more than 110 atoms. Across three boundary-pinning conditions, the measured response maps show the same qualitative organization as the calculated phase diagram. These results establish programmable Rydberg-atom arrays as tools for investigating frustrated, material-motivated Hamiltonians and predict $2/5$ and $3/5$ filling ($\pm1/5$ magnetization) plateaus and a quantum-dimer regime that could be sought using a transverse field in ErB$_4$ or strain in TmB$_4$.

\section*{Model}

\begin{figure*}[htb]
\centering
\includegraphics[width=1.0\textwidth]{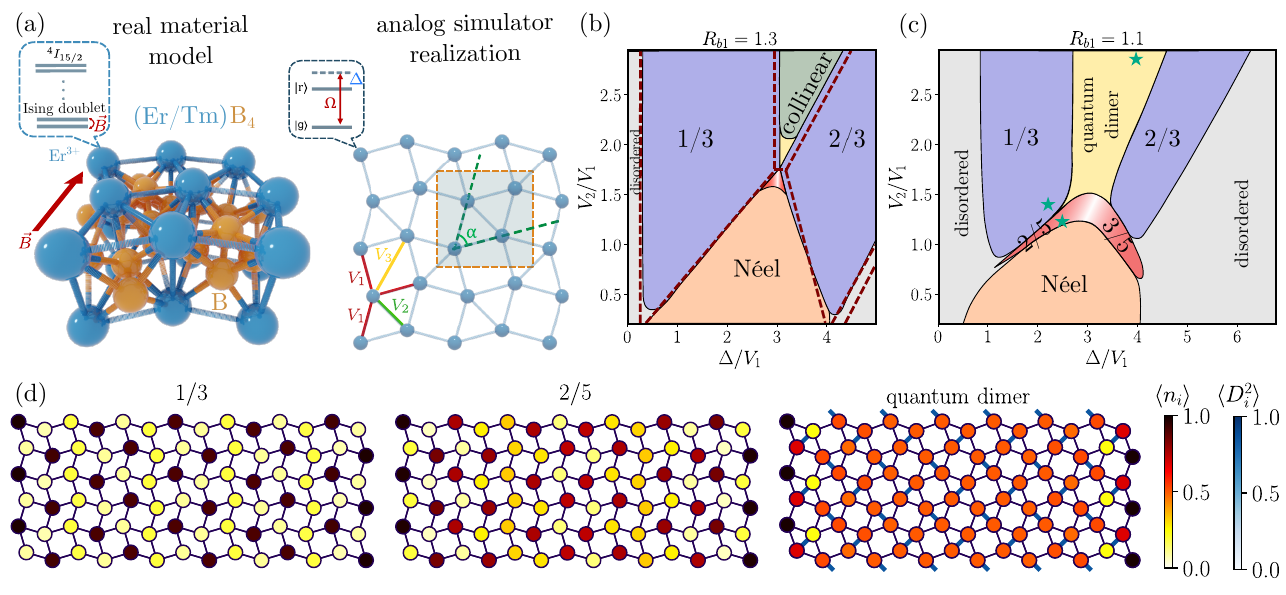}
\caption{\textbf{Quantum Shastry-Sutherland model motivated by rare-earth tetraborides and its Rydberg phase diagram.} (a) (left) Crystal structure of the rare-earth tetraboride family, which includes the Ising-like materials TmB$_4$ and ErB$_4$~\cite{Schafer1976}. The rare-earth ions form Shastry-Sutherland layers~\cite{Shastry1981} and are surrounded by boron cages. In ErB$_4$, the $^4I_{15/2}$ multiplet is split by the crystal-field environment into Kramers doublets, with the lowest-energy doublet forming the effective Ising degree of freedom. Longitudinal and transverse magnetic fields couple anisotropically to this doublet, while interactions between the rare-earth moments generate effective in-plane Ising couplings. (right) The corresponding effective spin model is mapped onto a programmable neutral-atom analog simulator~\cite{aquila:2023}, with the Ising states represented by the atomic ground and Rydberg states and geometrically tunable van der Waals interactions implementing the Ising couplings. The detuning and Rabi frequency control the effective longitudinal and transverse fields, respectively. (b)--(c) Ground-state phase-diagram cuts of the Rydberg Shastry-Sutherland model in Eq.~\ref{eq:RydHam}, obtained using density matrix renormalization group calculations~\cite{White1992,itensor} on a $14\times6$ cylinder (see Methods). At $R_{b1}=1.3$, quantum fluctuations are weak, and the resulting phase boundaries lie close to those of the exact classical model (red dashed lines)~\cite{Dublenych2013}. Increasing the transverse-field strength by decreasing $R_{b1}$ to $1.1$ produces extended $2/5$ and $3/5$ filling plateau regions and a quantum-dimer regime. (d) Local excitation densities at three representative points marked by green stars in (c), including the largely classical $1/3$ phase and the two regimes stabilized at finite transverse field. Because the excitation density is nearly uniform in the quantum-dimer regime, the local bond-dimerization diagnostic is shown instead.
}
\label{fig:Fig1_DMRG}
\end{figure*}

Rare-earth tetraborides crystallize in the tetragonal $P4/mbm$ structure~\cite{Schafer1976,Etourneau1985}, shown in Fig.~\ref{fig:Fig1_DMRG}. Their magnetic degrees of freedom derive from the spin-orbit-coupled ground multiplets of the rare-earth ions: the non-Kramers $^3H_6$ multiplet of Tm$^{3+}$ and the Kramers $^4I_{15/2}$ multiplet of Er$^{3+}$. The crystal electric field (CEF) generated by the surrounding boron cages further splits these multiplets, with the orthorhombic site symmetry allowing strong magnetic anisotropy~\cite{Fazekas1999,Flury2025}. Experiments establish strong Ising anisotropy in both TmB$_4$ and ErB$_4$~\cite{Schafer1976,Etourneau1985,Yoshii_2006,Ye2017,Flury2025}. These observations motivate an effective low-energy description in terms of a Kramers Ising doublet for ErB$_4$ and a likely non-symmetry-protected Ising pseudodoublet for TmB$_4$. We use a symmetry-constrained CEF model to assess the magnetic-field response of these low-energy degrees of freedom; the operator expansion and doublet-projection procedure are given in Methods.

The magnetism in both materials is governed by boron-mediated exchange and is predominantly two-dimensional, with much weaker coupling between rare-earth planes~\cite{Etourneau1985,Wierschem2013,Wierschem2015}. Given the strong Ising anisotropy and the Shastry-Sutherland geometry of the rare-earth layers [Fig.~\ref{fig:Fig1_DMRG}(a)], their magnetism can be described by an effective Ising Shastry-Sutherland Hamiltonian~\cite{Dublenych2012,Dublenych2013,Flury2025}. In particular, the similar distances and exchange pathways between first and second rare-earth neighbors motivate a frustrated model with $J_1\sim J_2$. With increasing longitudinal field $B_z$, TmB$_4$ progresses from N\'{e}el order through a sequence of fractional plateaus to a half-magnetization ($1/4$ filling) plateau. In ErB$_4$, the zero-field columnar antiferromagnetic order gives way to a half-magnetization plateau. Extended classical Ising models can account for these ordered states and transition sequences~\cite{Dublenych2012,Dublenych2013,Dublenych2014}, although stabilizing plateaus beyond the $1/3$ plateau of the minimal $J_1$--$J_2$ model requires longer-range Ising interactions. These include a third-neighbor coupling $J_3\ll J_1,J_2$ that lifts degeneracies of the $J_1$--$J_2$ model. Such longer-range interactions are naturally expected in metallic rare-earth compounds such as TmB$_4$ and ErB$_4$ because their magnetic exchange has Ruderman--Kittel--Kasuya--Yosida character~\cite{Ruderman1954}. These considerations motivate the minimal effective classical Hamiltonian used here:
\begin{align}\label{eq:Hcl}
H_{\mathrm{cl}}={}&J_1\sum_{\langle i,j\rangle}Z_iZ_j
+J_2\sum_{\langle\!\langle i,j\rangle\!\rangle}Z_iZ_j\\
&+J_3\sum_{\langle\!\langle\!\langle i,j\rangle\!\rangle\!\rangle}Z_iZ_j
-\frac{1}{2}g_z\mu_B B_z\sum_i Z_i,\notag
\end{align}
where the three bond sums are defined in Fig.~\ref{fig:Fig1_DMRG}(a), and the longitudinal field tunes the system between the ordered states.

The classical Shastry-Sutherland Ising model has an exactly characterized ground-state phase diagram~\cite{Dublenych2012,Dublenych2013,Dublenych2014,Huo2013}. Related anisotropic quantum extensions have also been studied~\cite{Meng2008}, and the classical $J_1$--$J_2$ phase structure has been reproduced on quantum-annealing hardware~\cite{Kairys2020}. The extensive degeneracies and field-induced plateaus of the classical model make Ising rare-earth tetraborides promising systems in which to investigate interplay between quantum effects and frustration. Here, we focus on ErB$_4$, where a transverse magnetic field may provide the required tuning. Because the ground state is a Kramers doublet, a transverse field can couple linearly to the effective spin when the transverse magnetic-dipole matrix elements are nonzero.  To assess the possible magnitude of this coupling, we constrain the CEF model using the measured saturation moment, $\mu=8.8(9)\,\mu_B$, and excitation gap, $\Delta_{\mathrm{CEF}}=8.2(1.0)\,\mathrm{meV}$~\cite{Flury2025}. These measurements do not directly determine the transverse $g$ factors; instead, they allow us to test whether appreciable transverse matrix elements are compatible with the known longitudinal moment and CEF gap. Starting from the published CEF description, we include additional symmetry-allowed higher-rank terms and diagonalize the resulting Hamiltonian while retaining both observables within their experimental uncertainty ranges (see Methods). Projecting the magnetic moment onto the resulting ground-state doublet yields an admissible parameter set with principal-axis $g$ factors $g_{\perp,1}=0.9$, $g_{\perp,2}=0.3$, and $g_z=16.1$, giving the transverse-field terms:\begin{equation}\label{eq:Htr}
H_{\perp}=\frac{1}{2}g_{\perp,1}\mu_B B_{\perp,1}\sum_i X_i
-\frac{1}{2}g_{\perp,2}\mu_B B_{\perp,2}\sum_i Y_i.
\end{equation}

Although these $g$ factors remain strongly Ising anisotropic, our calculation shows that the existing experimental constraints are compatible with a transverse $g$ factor of order unity. Estimates from the longitudinal transition fields and N\'eel temperature give $J_1\sim J_2\sim1\,\mathrm{meV}$. For the parameter set constructed above with $g_{\perp,1}=0.9$, an in-plane field of $10\,\mathrm{T}$ corresponds to a transverse coupling $\frac{1}{2}g_{\perp,1}\mu_B B_{\perp,1}\approx0.26\,\mathrm{meV}$. Thus, if transverse matrix elements of this magnitude are realized in ErB$_4$, laboratory-scale in-plane fields could access a regime in which quantum fluctuations appreciably compete with the Ising interactions.  In TmB$_4$, the non-Kramers pseudodoublet has only a quadratic leading response to an in-plane magnetic field, whereas in-plane strain can couple linearly. The required strain is difficult to estimate without first-principles modeling of TmB$_4$. These two tuning routes motivate quantum simulation as a means of identifying experimentally testable signatures of quantum frustration.

The transverse-field Ising model can be implemented directly with a Rydberg-atom analog quantum simulator. On the Aquila processor used here~\cite{aquila:2023}, atoms are trapped in programmable optical-tweezer arrays, with attainable system sizes exceeding 200 atoms~\cite{Ebadi2021}. Each atom is coherently driven between its ground state $\ket{g_i}$ and a highly excited Rydberg state $\ket{r_i}$ with Rabi frequency $\Omega$, detuning $\Delta$, and drive phase $\phi$. The resulting Hamiltonian is
\begin{align}\label{eq:RydHam}
H_{\mathrm{Ryd}}=&\frac{\Omega}{2}\sum_i
\left(e^{i\phi}\ket{g_i}\bra{r_i}+\mathrm{H.c.}\right)
-\Delta\sum_i n_i \\
& +\sum_{i<j}V_{ij}n_i n_j, \notag
\end{align}
where $n_i=\ketbra{r_i}{r_i}$ and $V_{ij}=C_6/|\mathbf{r}_i-\mathbf{r}_j|^6$ is the repulsive van der Waals interaction between two Rydberg excitations.

To make the material mapping explicit, we define the Pauli operators by $Z_i=\ketbra{r_i}{r_i}-\ketbra{g_i}{g_i}$, so that
$n_i=(1+Z_i)/2$.
For $\phi=0$, substituting this relation into Eq.~\ref{eq:RydHam} gives, up to an additive constant,
\begin{align}
H_{\mathrm{Ryd}}
=& \frac{\Omega}{2}\sum_iX_i
+\sum_{i<j}\frac{V_{ij}}{4}Z_iZ_j \\
& +\frac{1}{2}\sum_i\left(-\Delta+\frac{1}{2}\sum_{j\ne i}V_{ij}\right)Z_i. \notag
\end{align}
Comparison with Eqs.~\ref{eq:Hcl} and \ref{eq:Htr} therefore gives
$J_{ij}=V_{ij}/4$, 
$g_{\perp,1}\mu_BB_{\perp,1}=\Omega$, 
$g_z\mu_BB_{z,i}=\Delta-\frac{1}{2}\sum_{j\ne i}V_{ij}$.
The effective longitudinal field is uniform in a translationally invariant bulk system but acquires site-dependent boundary shifts in a finite open array. The mapping converts between filling and magnetization plateau conventions according to $M/M_\mathrm{sat}=\langle Z\rangle=2\langle n\rangle-1$, which we use interchangeably throughout the work.

The atomic geometry controls the interaction ratios and thereby provides access to the frustration axis of the phase diagram. In particular, changing the lattice skew angle $\alpha$ [Fig.~\ref{fig:Fig1_DMRG}(a)] tunes $V_2/V_1$ over a broad range around unity while preserving the Shastry-Sutherland bond connectivity. For the geometry considered here, $V_3=V_1/8$, while more distant interactions are suppressed by the $1/r^6$ decay. The detuning controls the effective longitudinal field, and the Rabi frequency controls the transverse field. We express the latter relative to the interaction scale using the dimensionless blockade radius
\begin{equation}
R_{b1}=\frac{R_b}{R_1}
=\left(\frac{V_1}{\Omega}\right)^{1/6},
\end{equation}
where $R_b=(C_6/\Omega)^{1/6}$ is the blockade radius and $R_1$ is the first-neighbor separation.

\section*{Results}

\subsection*{Theoretical simulations}

We determine the ground-state phase diagram of the material-motivated Ising Shastry-Sutherland model using density matrix renormalization group (DMRG) calculations~\cite{White1992,White1993,itensor} (see Methods) and subsequently probe it on quantum hardware. The main calculations use $14\times6$ cylinders, which capture bulk ordering tendencies when the correlation length and ordering period fit within the finite geometry, but do not constitute a two-dimensional thermodynamic-limit extrapolation. Nevertheless, the simulations should correspond well to the expected bulk behavior, provided the unit cell of the magnetic phase in question is small. Phase diagrams at $R_{b1}=1.3$ and $1.1$ are shown in Fig.~\ref{fig:Fig1_DMRG}(b)--(c). Candidate phase boundaries are located by following ridges in the half-system von Neumann entanglement entropy across the sampled parameter grid (see Supplementary Information). The phase assignments on either side are then cross-checked using real-space density patterns and the dominant static structure-factor peaks. Because these calculations are performed on finite cylinders and a discrete parameter grid, the lines in Fig.~\ref{fig:Fig1_DMRG}(b)--(c) should be interpreted as numerical estimates of the phase boundaries rather than thermodynamic-limit critical points. The ordered regions correspond to magnetic order in the effective-spin language and density-wave order in the Rydberg-atom language.

At weak transverse field, $R_{b1}=1.3$ ($\Omega/V_1\approx0.2$), the phase diagram closely follows the exact classical $J_1$--$J_2$--$J_3$ phase diagram, whose boundaries are overlaid as red dashed lines~\cite{Dublenych2013}. The low-$V_2/V_1$ region supports N\'{e}el order at zero longitudinal field, whereas at larger $V_2/V_1$ the classical $V_1$--$V_2$ model hosts an extensively degenerate dimer manifold from which $V_3$ selects collinear order~\cite{Shastry1981,Dublenych2012,Dublenych2013}. Varying the detuning produces broad $1/3$ and $2/3$ filling plateaus and narrow classical $1/4$ and $3/4$ plateaus; the latter are absent at $R_{b1}=1.3$, consistent with their weaker stabilization by $V_3$. This organization provides a qualitative connection to rare-earth tetraborides. The larger-$V_2/V_1$ regime captures the progression in ErB$_4$ from columnar antiferromagnetic order to a half-magnetization plateau~\cite{Flury2025}, corresponding to a filling of $1/4$ or $3/4$. TmB$_4$ is qualitatively associated with the lower-$V_2/V_1$ regime and exhibits additional narrow $1/7$, $1/8$, and $1/9$ magnetization plateaus attributed to longer-range interactions~\cite{Siemensmeyer2008,Trinh2018,Dublenych2014}; these fine plateaus are not resolved in the finite-transverse-field phase diagrams studied here, as quantum fluctuations overwhelm any stabilizing effects of $V_k, \, k>3$.

\begin{figure}[t]
\centering
\includegraphics[width=1.0\columnwidth]{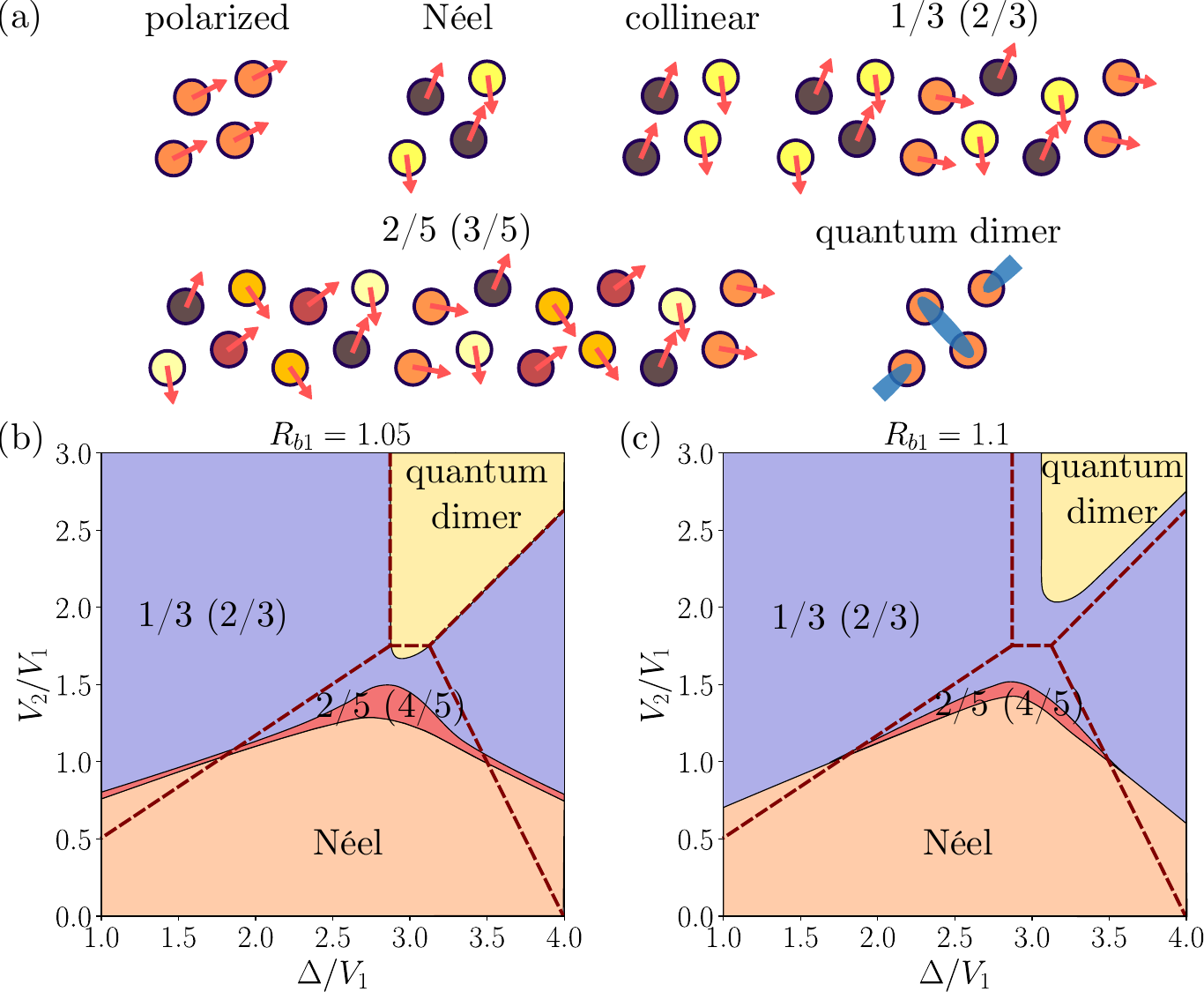}
\caption{\textbf{Variational modeling of the transverse-field Shastry-Sutherland phase diagram.} (a) The states identified numerically are modeled using minimal variational Ans\" atze. Single-site Ans\" atze are shown within the unit cell of each ordered state, with colors and arrow directions denoting the independent parameters $\theta_i$. The quantum-dimer regime is described by an entangled two-site Ansatz that admixes the one-excitation dimer states with zero- and two-excitation configurations. (b)--(c) Phase diagrams obtained by variationally minimizing the energy within the Ansatz spaces shown in (a) at two transverse-field strengths (see Methods, smaller $R_{b1}$ corresponds to a stronger transverse field). The results reproduce the qualitative DMRG trends in Fig.~\ref{fig:Fig1_DMRG}(b)--(c), with a stronger transverse field enlarging the $2/5$-ordered and quantum-dimer regions.}
\label{fig:Fig2_Ansatze}
\end{figure}

Beyond suppressing the narrow classical plateaus and shifting phase boundaries, a finite transverse field stabilizes two new regimes near the strongly frustrated central boundary: the $2/5$ and $3/5$ filling plateaus and the quantum-dimer regime. Their characteristic density and bond patterns are shown in Fig.~\ref{fig:Fig1_DMRG}(d). The fractional labels refer to the fillings of the corresponding ideal classical patterns; at finite transverse field, the mean densities deviate from these values, and the ordered states are identified by their symmetry-breaking patterns and order parameters (see Supplementary Information). At $R_{b1}=1.3$, these regimes occupy narrow parameter ranges, but they expand substantially at $R_{b1}=1.1$ ($\Omega/V_1\approx0.56$) and dominate the most frustrated region of the phase diagram [Fig.~\ref{fig:Fig1_DMRG}(c)]. By contrast, the classically stabilized N\'{e}el, $1/3$-filling, and collinear phases shrink as the transverse field increases. The fluctuation-stabilized regimes therefore emerge most prominently at intermediate transverse-field strengths before giving way to a strongly driven paramagnet. Possible critical behaviors at the intervening transitions are analyzed using symmetry-constrained Landau theories in the Supplementary Information.

 We now describe and model how the interplay between quantum fluctuations and frustration stabilizes new phases. We first consider the $2/5$-filling phase, which lies between the $1/3$-filling and N\'{e}el phases. Its ideal classical pattern consists of four N\'{e}el-ordered columns followed by one empty column, repeated within a ten-site unit cell; the particle-hole-related $3/5$ pattern instead contains a filled column. At the classical $1/3$--N\'{e}el boundary, this pattern is exactly degenerate with the adjacent phases as part of an extensive manifold containing different arrangements of N\'{e}el-ordered and polarized stripes~\cite{Dublenych2013}. A finite transverse field lifts this degeneracy and stabilizes the next simplest (smallest unit cell) ordered excitation pattern --- $2/5$ filling phase over a finite parameter interval, providing an example of quantum order by disorder~\cite{Villain1980,Henley1987,Chandra1990,Chubukov1991,Chubukov1992}. The small unit cell ordered phase stabilization is typical for a quantum order-by-disorder mechanism, as coherent momentum-space excitation modes typically have lower average zero-point energy than more generic states in the classically degenerate manifold. The DMRG state exhibits substantial occupation of both classically empty and occupied sites, consistent with coherent quantum mixing. Moreover, the $2/5$ region persists across the cylinder lengths studied (see Supplementary Information), including geometries incommensurate with its unit cell, supporting its interpretation as a bulk ordering tendency rather than boundary-induced stabilization.

To model the order-by-disorder mechanism, we use a minimal variational description (see Methods). The polarized, N\'{e}el, collinear, $1/3$-filling, and $2/5$-filling states are approximated by symmetry-constrained product states,
\begin{equation}
|\Psi_A[\{\theta_i\}]\rangle
=\bigotimes_i\left(\cos\theta_i|r_i\rangle+\sin\theta_i|g_i\rangle\right),
\end{equation}
with the parameters constrained according to the unit cell of each state [Fig.~\ref{fig:Fig2_Ansatze}(a)]. In the classical limit, these Ans\" atze reproduce the relevant phase boundaries between the included candidate states. At finite $\Omega$, they also capture the main qualitative DMRG trends: a finite $2/5$-filling region opens along the classical $1/3$--N\'{e}el boundary, the boundary shifts, and the collinear phase is suppressed [Fig.~\ref{fig:Fig2_Ansatze}(b)--(c)]. At the same transverse-field strength, $R_{b1}=1.1$, the product-state calculation yields a narrower $2/5$ region than DMRG [compare Figs.~\ref{fig:Fig2_Ansatze}(c) and~\ref{fig:Fig1_DMRG}(c)]. Thus, on-site quantum coherence already captures the selection of this order, while its quantitative stability extends beyond the restricted product-state description. For the same reason, the Ansatz is not expected to capture the critical behavior at the phase transitions discussed separately in the Supplementary Information, as the product state picture can only recover the mean-field Landau theory results.

We next consider the quantum-dimer regime. In the classical $V_1$--$V_2$ model, the central large-$V_2/V_1$ region contains an extensively degenerate manifold of dimer configurations, from which the weak $V_3$ interaction selects collinear order. At finite transverse field, this region is instead replaced over an extended parameter range by a symmetry-preserving, short-range-entangled dimer state. An idealized representative is the Bell-dimer product state
\begin{equation}
|\Psi_{\mathrm{dimer}}\rangle
=\bigotimes_{(i,\langle i\rangle^2)}
\frac{|g_i r_{\langle i\rangle^2}\rangle
+|r_i g_{\langle i\rangle^2}\rangle}{\sqrt{2}},
\end{equation}
where the product includes each $V_2$ bond once and $\langle i\rangle^2$ denotes the unique site paired with $i$ on that bond. The local excitation density is uniform, with $\langle n_i\rangle=1/2$. We define the dimer-orientation operator
\begin{equation}
D_i=n_i-n_{\langle i\rangle^2},
\end{equation}
which takes values $\pm1$ for the classical configurations $|g_i r_{\langle i\rangle^2}\rangle$ and $|r_i g_{\langle i\rangle^2}\rangle$ and vanishes for the zero- and two-excitation configurations. In the quantum-dimer regime, $\langle D_i\rangle$ is nearly zero, while $\langle D_i^2\rangle$ is large and approximately uniform across the $V_2$ bonds [Fig.~\ref{fig:Fig1_DMRG}(d)], demonstrating a strong local one-excitation constraint. Correlations between dimer-orientation operators on different bonds decay rapidly, indicating the absence of long-range dimer order. The coherent quantum character of the regime is instead captured by enhanced $\langle X_iX_{\langle i\rangle^2}\rangle$ correlations on the $V_2$ bonds (see Supplementary Information). Because this state is adiabatically connected to the paramagnet, it does not constitute a distinct symmetry-breaking phase; nevertheless, its strong local dimer correlations and bond-selective transverse coherence define a physically distinct crossover regime.

The stabilization mechanism of the quantum-dimer regime can be understood as a competition between the classical energy selecting collinear order and the transverse-field energy gain within the dimer manifold. Classically, $V_3$ selects collinear order. At finite transverse field, the symmetric dimer state can lower its energy through virtual admixture of zero- and two-excitation configurations (see Methods). A simple second-order estimate compares an energy gain of order $\Omega^2/V_2$ per dimer with the collinear-selection energy of order $V_3$. Using $V_3=V_1/8$ gives
\begin{equation}
R_{b1}^{\mathrm{coll}}
\sim\left(\frac{V_2}{8V_1}\right)^{-1/12},
\end{equation}
or $R_{b1}^{\mathrm{coll}}\sim1.08$ at $V_2/V_1=3$. This estimate captures the relevant scale but underestimates the extent of the dimer regime in DMRG, indicating that higher-order processes and correlations are important. To incorporate local dimer entanglement and fluctuations of the competing collinear state, we supplement the product-state Ans\" atze with a two-site dimer Ansatz [Eq.~\ref{eq:dimer_Ansatz}] that coherently admixes the one-excitation dimer configurations with zero- and two-excitation components. Variational optimization qualitatively reproduces the replacement of the classical collinear region by the quantum-dimer regime as $R_{b1}$ decreases, as well as the growth of this regime with increasing transverse field. The two-site Ansatz therefore provides a simple microscopic account of the energy competition, while the DMRG results capture additional many-body correlations within this short-range-correlated, symmetry-preserving regime.

\subsection*{Experiments}

\begin{figure*}[t]
\centering
\includegraphics[width=\textwidth]{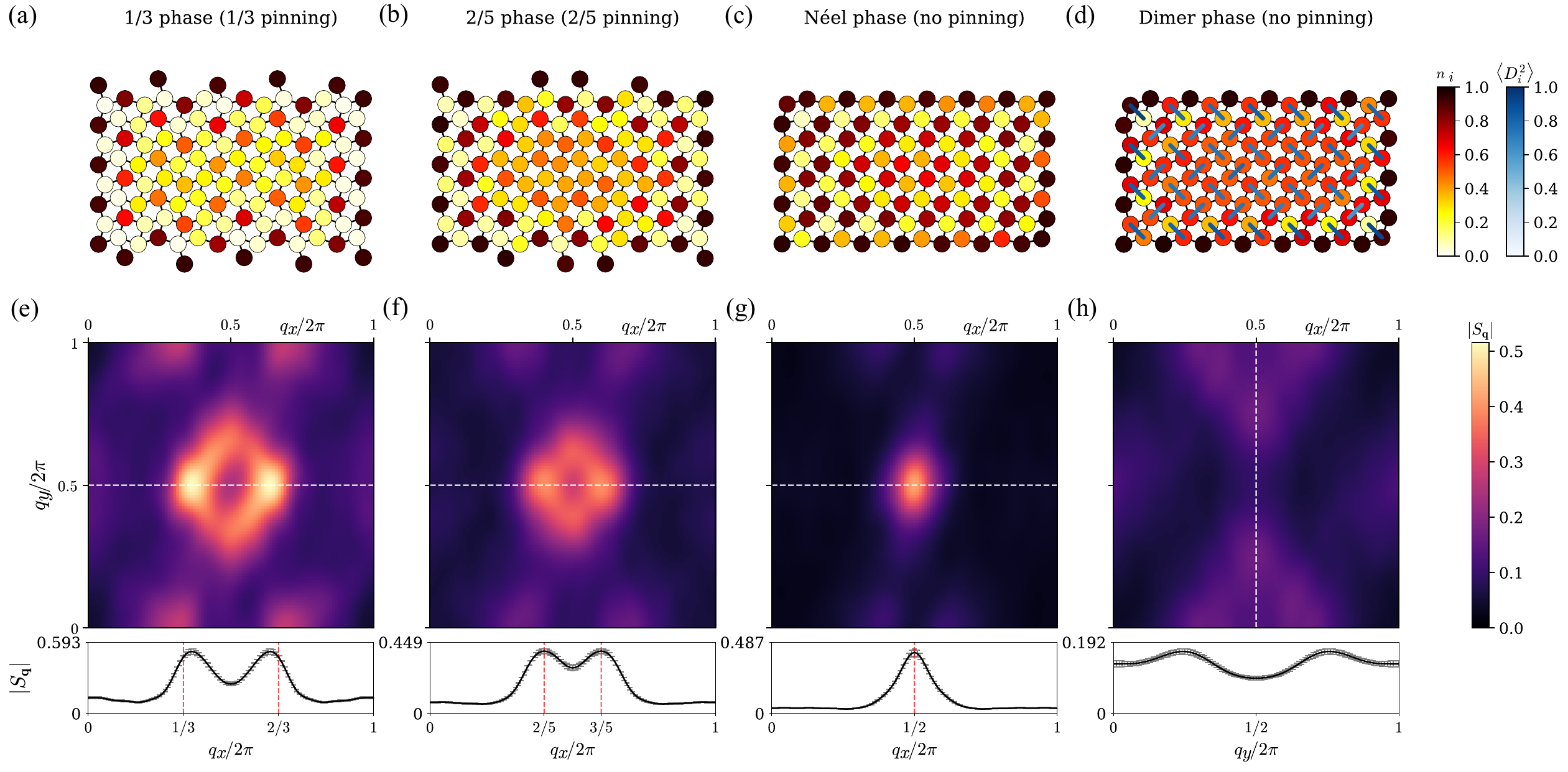}
\caption{\textbf{Representative experimental signatures of phases in the quantum Shastry-Sutherland lattice.} (a)--(d) Rydberg-density profiles, $\langle n_i\rangle$, of states prepared using detuning sweeps on the Aquila analog quantum simulator (see Methods)~\cite{aquila:2023}. Specifically, (a) shows the $1/3$-filling ordered state prepared with $1/3$-favoring boundary pinning, (b) shows the $2/5$-filling ordered state prepared with $2/5$-favoring boundary pinning, and (c) shows the N\'{e}el state prepared without auxiliary pinning atoms, for which the native open boundary favors N\'{e}el order. (d) shows experimental signatures at a parameter point associated with the quantum-dimer regime, also without auxiliary pinning atoms, characterized by the local dimer-occupation diagnostic $\langle D_i^2\rangle$. (e)--(h) Corresponding experimentally extracted static structure factors $S_{\mathbf q}$ (see Eq.~\ref{eq:SSF}) for the states shown in (a)--(d), together with representative cuts at $q_y=\pi$ [$q_x=\pi$ in (h)]. Error bars denote the standard error of the mean estimated from the experimental shots (see Methods). The ordered states exhibit peaks at the expected ordering wavevectors, whereas the density structure factor in the dimer regime shows only diffuse features, consistent with the absence of long-range density order. The parameter points corresponding to these representative states are marked in Fig.~\ref{fig:Fig4_PDs}.}
\label{fig:Fig3_phases}
\end{figure*}

We now turn to the experimental realization of the Shastry-Sutherland Rydberg Hamiltonian and the measurement of signatures across its phase diagram. The experiments were performed on the Aquila analog quantum simulator~\cite{aquila:2023} by sorting single atoms into programmable tweezer arrays with Shastry-Sutherland geometry. For each experimental cut, the nearest-neighbor lattice constant was chosen to fix the dimensionless blockade radius $R_{b1}=(V_1/\Omega)^{1/6}=1.1$. The frustration ratio $V_2/V_1$ was then tuned geometrically by changing the lattice skew angle $\alpha$ [Fig.~\ref{fig:Fig1_DMRG}(a)], which controls the length of the shorter Shastry-Sutherland bonds at fixed $V_1$. To access the strongly frustrated regime while preserving broad tunability of $V_2/V_1$, we modified the reservoir packing and row-specific trap allocation used in Aquila's atom-sorting procedure, enabling reduced inter-row spacing and taller arrays at the cost of increased pattern width (see Methods). This allowed us to use a bulk lattice of size $N_x=8$ and $N_y=14$, the largest compatible with these geometry and trap-budget constraints.

State preparation was carried out using finite-duration sweeps of the global detuning under a constant Rabi drive (see Methods)~\cite{Keesling2019,Ebadi2021}, starting from the trivial all-zero state and passing through a strongly quantum-mixed regime near zero detuning. Since the finite sweep duration and potentially small many-body gaps preclude a general guarantee of adiabatic ground-state preparation, we interpret the observed states as experimental signatures of the corresponding numerically identified regimes. 

Because the experiments are performed on finite open arrays, boundary effects remain important~\cite{Kalinowski2022,Rourke2023,Hirsbrunner2025}: atoms near the edge have reduced coordination and are consequently more readily excited, which can prevent them from following the bulk ordering pattern. To control and characterize these boundary effects, we introduced auxiliary atoms at selected boundary positions chosen to favor the ordering pattern under study [see Fig.~\ref{fig:Fig3_phases}(a)--(b)]. We repeated the measurements for three boundary conditions at fixed bulk size, designed respectively to favor the $1/3$-filling order, the $2/5$-filling order, and the N\'{e}el order, with the last requiring no auxiliary pinning atoms. The resulting measurements should therefore not be interpreted as a direct thermodynamic-limit phase diagram. Instead, the three boundary conditions provide controlled finite-size biases toward different ordering patterns, allowing us to test whether the corresponding diagnostic responses remain localized to similar regions of Hamiltonian parameter space. Future experiments with smaller boundary-to-bulk ratios could further reduce these boundary effects.

Representative experimental signatures of the four principal regimes are shown in Fig.~\ref{fig:Fig3_phases}. Two correspond directly to the dominant classical orders. The $1/3$-filling order is already apparent in the spatial Rydberg density, where the measured pattern exhibits the expected alternation of two N\'{e}el-like columns and one nearly unexcited column, with the aspect ratio and boundary pinning selecting a dominant configuration. Likewise, the N\'{e}el order is readily identified by its staggered density pattern at low $V_2/V_1$. The two regimes stabilized by quantum fluctuations require more discriminating diagnostics. In the quantum-dimer regime, the average density remains nearly uniform and close to $1/2$ in the bulk, while the dimer observable $\langle D_i^2\rangle$ is large and nearly uniform on the bulk $V_2$ bonds. These observations are consistent with strong local one-excitation character on the dimers and no resolved long-range density modulation, although $\langle D_i^2\rangle$ alone does not probe quantum coherence within each dimer. By contrast, the $2/5$-filling order is considerably harder to distinguish from the density profile alone: finite system size and competing nearly degenerate configurations reduce the contrast of the ideal ten-site-unit-cell pattern in the averaged density, making the density profile suggestive but not a reliable discriminator in the experimentally relevant regime.

\begin{figure*}[t]
\centering
\includegraphics[width=\textwidth]{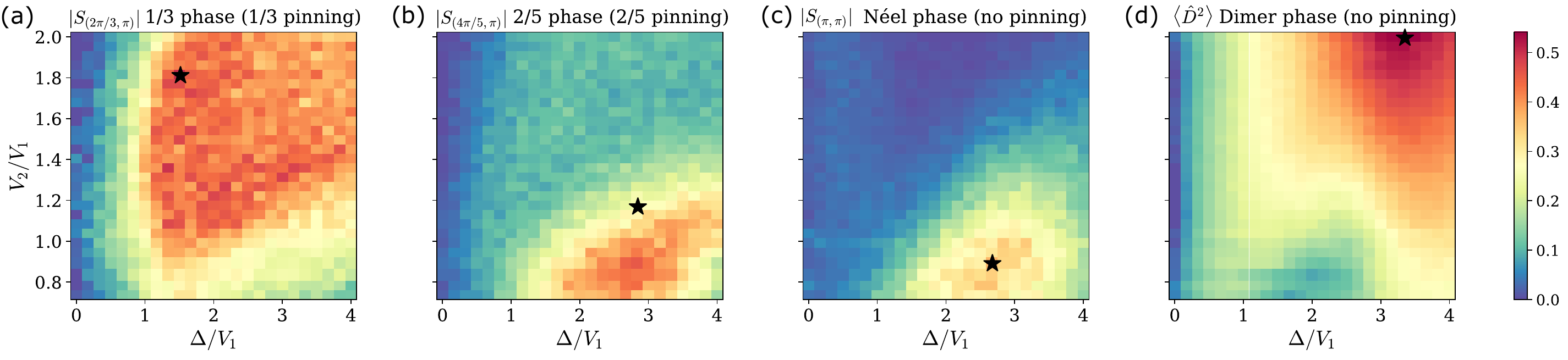}
\caption{\textbf{Experimental order-parameter response maps of the Rydberg Shastry-Sutherland lattice.} Experimentally measured structure-factor amplitudes, $\left|S_{\mathbf q} \right|$, at the dominant ordering wavevectors for the density-ordered states and the average local dimer occupation, $\langle D^2\rangle=(1/\mathcal{N})\sum_i\langle D_i^2\rangle$, for the dimer regime, plotted as functions of $\Delta/V_1$ and $V_2/V_1$ at fixed $R_{b1}=1.1$. Here, $\mathcal{N}$ is the number of bulk $V_2$ bonds included in the average. (a) $1/3$-filling structure-factor response under $1/3$-favoring boundary pinning. (b) $2/5$-filling structure-factor response under $2/5$-favoring boundary pinning. (c) N\'{e}el structure-factor response and (d) local dimer-occupation response in the native unpinned geometry, whose open boundaries favor N\'{e}el order. Stars mark the representative experimental points shown in Fig.~\ref{fig:Fig3_phases}. The four diagnostics are enhanced in regions that agree qualitatively with the numerical phase diagram in Fig.~\ref{fig:Fig1_DMRG}. Their contrast and apparent extent depend on the boundary condition, reflecting the finite open geometry (see Supplementary Information).}
\label{fig:Fig4_PDs}
\end{figure*}

The static structure factor provides a more sensitive diagnostic of spatial density-wave correlations:
\begin{equation}\label{eq:SSF}
    S_{\mathbf q}=\frac{1}{\mathcal{N}_Q}\sum_{i,j}e^{-i\mathbf q\cdot(\mathbf r'_i-\mathbf r'_j)}
    \left\langle \left(n_i-\frac{1}{2}\right)\left(n_j-\frac{1}{2}\right)\right\rangle,
\end{equation}
where $\mathbf r'_i$ denotes the position of atom $i$ on the topologically equivalent square lattice (see Supplementary Information), $\mathbf q$ is the corresponding reciprocal-lattice vector, and the normalization $\mathcal{N}_Q$ is chosen such that $S_{\mathbf Q}=1$ for a perfect classical representative of a given ordered phase. For the dimer-regime data, we use the normalization defined by the perfect classical collinear state. This provides a reference scale for comparing the diffuse dimer-regime signal with perfect collinear order; it does not assume collinear order in the prepared state. Compared with the average real-space density, $S_{\mathbf q}$ remains a robust discriminator when the measured state contains admixtures of symmetry-related configurations, such as different translations or the two orientational variants of the $2/5$-filling order. As shown in Fig.~\ref{fig:Fig3_phases}(e)--(h), the experimentally extracted structure factors exhibit peaks at the wavevectors expected from numerics: $(\pi,\pi)$ for N\'{e}el order; $(\pm2\pi/3,\pi)$ and $(\pi,\pm2\pi/3)$ for the $1/3$-filling order, with a strong preference for the horizontal orientation; and $(\pm4\pi/5,\pi)$ and $(\pi,\pm4\pi/5)$ for the $2/5$-filling order. By contrast, the density structure factor in the quantum-dimer regime shows no sharp peak and only diffuse weight broadly concentrated near $(\pi,0)$ and $(0,\pi)$. Together with the large local dimer occupation $\langle D_i^2\rangle$, this is consistent with short-range density correlations associated with local dimer formation and the absence of resolved long-range density order. The finite system size limits the experimental momentum resolution, while finite sampling limits the statistical precision. Consequently, the proximity of the N\'{e}el ordering vector to the nearby $2/5$ peaks makes the latter particularly difficult to resolve, as seen in the cuts at fixed $q_y=\pi$. The peak magnitudes of the ordered phases remain well below the ideal classical value of $1$, consistent with finite correlation lengths in the prepared states, experimental noise, and redistribution of spectral weight between orientational variants. Nevertheless, the ordered phases exhibit substantially stronger peak responses than the quantum-dimer regime, providing a clear contrast between density-ordered and dimer-regime responses within the present finite-size, finite-time, and noise limitations.

We use the static structure factors for the ordered phases and $\langle D^2\rangle$ for the dimer regime to construct the experimental response map in the relevant frustrated region of $(\Delta/V_1,V_2/V_1)$ parameter space, as summarized in Fig.~\ref{fig:Fig4_PDs}.  Overall, the experimental responses occur in regions broadly corresponding to the numerical phase diagrams, although differences in their extents likely reflect open-boundary and commensurability effects, as well as finite-time state preparation. In the main text, we show the ``native pinning'' configuration for each regime, namely the boundary condition under which its diagnostic is most clearly expressed. Under these conditions, the $1/3$-filling and N\'{e}el regions are the easiest to delineate, while the $2/5$-filling response appears in the same central frustrated region where it is predicted numerically. Importantly, the corresponding diagnostic responses are not exclusive to their native pinning conditions. As shown in the Supplementary Information, signatures of all four regimes remain visible across the three boundary conditions and remain localized to similar regions of parameter space. The auxiliary atoms are the additional boundary sites shown in Fig.~\ref{fig:Fig3_phases}(a)--(b); their interactions modify the local excitation energies near the array edges and bias compatible ordering patterns. The persistence of enhanced responses in similar parameter regions under different boundary choices is consistent with ordering tendencies governed by the bulk Hamiltonian. Boundary pinning therefore acts as a symmetry-selective field that enhances the response toward a preferred order and improves its contrast in the finite open system.

This cross-pinning comparison is especially important for interpreting the fluctuation-stabilized regimes. As a benchmark, the $1/3$-filling order remains robust under alternate boundary conditions, although its response occupies the largest region under its native pinning. The $2/5$-filling response follows the same qualitative trend but is more difficult to isolate because its ordering wavevector lies close to that of the N\'{e}el order relative to the finite experimental momentum resolution. The $2/5$ response is strongest in the central frustrated region near the N\'{e}el--$1/3$ boundary, where the numerical phase diagram also places the $2/5$ regime, although overlap between the two structure-factor responses limits their clean separation.  This overlap need not imply phase coexistence, since finite-size broadening already mixes the responses at nearby ordering wavevectors. Conversely, the present measurements do not exclude contributions from mixed-domain or metastable configurations produced during preparation. Changing the boundary condition selectively enhances the expected $2/5$ response, but the response remains visible in the native unpinned geometry (see Supplementary Information). The local dimer-occupation diagnostic $\langle D^2\rangle$ is comparatively insensitive to the boundary condition, consistent with the quantum-dimer regime being a short-range correlated regime rather than a long-range symmetry-breaking phase.  Overall, the experimental response maps are qualitatively consistent with the numerical phase organization, while finite-size, boundary, and state-preparation effects prevent a precise experimental determination of the thermodynamic phase boundaries.

\section*{Discussion and Outlook}

 In this work, we establish a direct route from the frustrated Ising physics of rare-earth tetraborides to a programmable neutral-atom simulator. Starting from experimentally constrained crystal-field descriptions, we show that appreciable transverse magnetic couplings are compatible with, although not directly established by, existing measurements of ErB$_4$ and identify strain as a route to transverse pseudospin dynamics in TmB$_4$. We then map this material-motivated Hamiltonian onto a Rydberg realization with independent control over longitudinal field, transverse-field scale, and the ratio of competing couplings, thereby opening controlled cuts through a genuinely three-dimensional phase diagram.  Our results show that a finite transverse field can qualitatively reorganize the classically frustrated Ising Shastry-Sutherland model rather than merely shift its classical phase boundaries. Ground-state DMRG calculations show that quantum fluctuations select $2/5$- and $3/5$-filling order from a classically degenerate boundary manifold and stabilize it over finite parameter intervals absent from the classical reference phase diagram. They also stabilize a symmetry-preserving quantum-dimer regime in place of $V_3$-selected collinear order. Minimal variational Ans\" atze trace both effects to the way transverse dynamics lift the underlying classical degeneracies. Experiments in finite arrays containing more than 110 atoms reveal density, structure-factor, and local dimer-occupation signatures that follow the qualitative organization of the calculated phase diagram across three boundary conditions. Although these finite-system measurements do not by themselves establish thermodynamic phases or transverse dimer coherence, the consistency across diagnostics and boundary conditions supports the interpretation that the observed responses reflect intrinsic ordering tendencies of the Hamiltonian rather than patterns imposed solely by the boundaries. Together, these results establish a controlled connection between programmable Rydberg arrays and the frustrated Ising physics relevant to rare-earth tetraborides, providing a basis for concrete material predictions.

These results motivate two concrete experimental tests in Ising Shastry-Sutherland materials in which transverse pseudospin dynamics can be introduced through an in-plane magnetic field in ErB$_4$ or strain in TmB$_4$. The first is a fluctuation-stabilized magnetization plateau: the phases adiabatically connected to $2/5$ and $3/5$ filling plateaus correspond to magnetizations $-1/5$ and $+1/5$, respectively, arising near the boundaries between the N\'{e}el state and the conventional $1/3$-magnetization plateau. The calculations also predict a nonmonotonic dependence on the transverse coupling: the plateau initially broadens as the transverse term is introduced and then contracts once the transverse field becomes strong enough to suppress longitudinal order. This nonmonotonic behavior reflects the competing roles of transverse fluctuations -- they first lift the classical degeneracy in favor of $2/5$- and $3/5$-ordered states, but ultimately destabilize longitudinal order itself. Our material-constrained CEF estimates suggest that transverse effects in ErB$_4$ could become appreciable at fields of order $10\,\mathrm{T}$, although the field required to reach the plateau depends sensitively on the actual transverse $g$ factor and exchange parameters~\cite{Flury2025}. If realized, these plateau phases can be validated by their ordering wavevectors in neutron diffraction. The second directly testable prediction is the appearance of a non-symmetry-breaking quantum dimer regime in the transverse field, replacing the classically expected collinear (columnar antiferromagnetic) phase. The dimer regime should instead exhibit diffuse magnetic scattering associated with short-range correlations rather than sharp ordering peaks. Polarization-resolved measurements and inelastic neutron spectroscopy could further test the predicted transverse correlations and dimer dynamics~\cite{Bramwell2014,Fogh2024}. Together, these plateau and dimer signatures provide stringent tests of whether the effective Shastry-Sutherland Ising description remains predictive when transverse quantum dynamics are introduced. 

The most immediate priorities on the simulator-side are to strengthen the equilibrium identification of these regimes and directly establish the coherent character of the quantum-dimer regime. Larger arrays with smaller boundary-to-bulk ratios would enable more systematic finite-size tests of the ordered regions, while optimized state-preparation protocols could reduce dynamical defects. The key additional dimer observable is the bond-selective transverse coherence $\langle X_iX_{\langle i\rangle^2}\rangle$. Together with large $\langle D_i^2\rangle$ and short-ranged dimer-orientation correlations, this observable would distinguish coherent resonance between the two local dimer configurations from an incoherent mixture with the same local occupation. It is not accessible from the native $Z$-basis measurements used here, but basis-rotation and analog-digital measurement protocols provide a concrete route toward measuring it in future Rydberg-array experiments~\cite{Semeghini2021,geim2026engineeringquantumcriticalitydynamics}. Such measurements would close the central gap between the present density-based experimental signatures and the DMRG characterization of the quantum-dimer regime.

More broadly, the material-to-simulator mapping provides a practical route to exploring both equilibrium phase diagrams and the dynamics of frustrated phase formation. Independent control of the interaction ratio, longitudinal field, and transverse coupling allows broad parameter scans within a single experimental setting, helping identify promising regimes and discriminating observables before undertaking targeted field- or strain-dependent measurements in the materials. Recent experiments on new Er$^{3+}$ Shastry-Sutherland material, Er$_2$Be$_2$GeO$_7$, sharpen this opportunity: a weak orthorhombic distortion was shown to qualitatively reorganize the plateau sequence, while quantum fluctuations were implicated in lifting residual plateau degeneracies, suggesting that the independent geometric and transverse-field controls available here could disentangle these effects in a controlled setting~\cite{Yadav2025}. Controlled ramps and quenches across the ordered-phase boundaries and through the dimer regime could reveal how competing orders nucleate, form domains, coarsen, or become trapped in long-lived nonequilibrium states. Related Rydberg-array experiments have already demonstrated access to nonergodic dynamics, quantum coarsening, and prethermal behavior~\cite{Bernien2017,Manovitz:2025,Darbha2025,Balewski2025,Hudomal2025}. Testing corresponding phenomena in materials could draw on high-field neutron spectroscopy and time-resolved X-ray probes, including resonant inelastic scattering and absorption, which require specialized facilities and microscopic modeling to connect measured signals to the underlying dynamics~\cite{Bramwell2014,Fogh2024,deGroot2024,Mitrano2024,Jost2025}. The simulator can supply a controlled reference for interpreting these measurements. This capability is particularly valuable for frustrated quantum magnets, whose real-time dynamics remain challenging for classical methods because of dynamical sign problems in Monte Carlo approaches and entanglement growth in tensor-network simulations~\cite{Henelius2000, Daley2022, Burkard2026}. Programmable simulators therefore provide a natural complement to microscopic modeling and materials measurements~\cite{Georgescu2014,Daley2022}. By making frustration and transverse dynamics separately tunable, they could distinguish consequences intrinsic to the Shastry-Sutherland Hamiltonian from those arising from material-specific physics—and translate that distinction into targeted tests in quantum magnets.

\begin{acknowledgments}
We acknowledge fruitful discussions with P. Lotshaw and P. Lougovski. This work was supported by the U.S.~Department of Energy, Office of Science, National Quantum Information Science Research Centers, Quantum Science Center. We also acknowledge support by AWS Cloud Credit for Research. Purdue team would like to acknowledge the support from U.S.~Army Research Office (ARO) under grant number W911NF2510290 and Purdue’s Elmore ECE Emerging Frontiers Center: The Crossroads of Quantum and AI.
\end{acknowledgments}

\section*{Methods}
\subsection*{Numerical methods}

\subsubsection*{Crystal field modeling}

We describe the crystal electric field within the spin-orbit-coupled ground-state $J$ multiplet using the Stevens-operator expansion~\cite{Stevens1952,Fazekas1999},
\begin{equation}\label{eq:CEF}
    H_{\mathrm{CEF}} =\sum_{l=2,4,6}\sum_{m=0,2,\ldots,l}B_l^m O_l^m,
\end{equation}
where $O_l^m$ are the Stevens operators allowed by the orthorhombic site symmetry and $B_l^m$ are their coefficients. The operators are polynomials in the angular-momentum components acting within the fixed-$J$ multiplet.

For ErB$_4$, we diagonalize this Hamiltonian in the $J=15/2$ manifold of Er$^{3+}$. We fix the rank-2 coefficients to $B_2^0=0.14\,\mathrm{meV}$ and $B_2^2=-0.26\,\mathrm{meV}$, as estimated in Ref.~\cite{Flury2025}. The sizable $B_2^2$ term strongly mixes the $m_J$ components of the ground-state doublet, facilitating transverse matrix elements. The remaining symmetry-allowed coefficients are varied in pairs to seek appreciable transverse coupling, with each varied coefficient restricted to $|B_l^m|<10^{-4}\,\mathrm{meV}$. We retain parameter sets for which the saturation moment and the gap from the ground-state doublet to the first excited CEF level remain within the measured ranges $\mu=8.8(9)\,\mu_B$ and $\Delta_{\mathrm{CEF}}=8.2(1.0)\,\mathrm{meV}$.

To determine the effective magnetic response, we project the angular-momentum operators onto the ground-state Kramers doublet. For an orthonormal doublet basis $\{\ket{+},\ket{-}\}$, the projected moment matrices have elements $(\widetilde J_\alpha)_{ab}=g_J\langle a|J_\alpha|b\rangle$, where $a,b\in\{+,-\}$, $\alpha\in\{x,y,z\}$, and $g_J=6/5$ for the $^4I_{15/2}$ multiplet. The effective $g$ tensor is $G_{\alpha\beta}=\mathrm{Tr}\left(\widetilde J_\alpha\sigma_\beta\right)$, where $\sigma_\beta$ are Pauli matrices in the doublet basis and the trace is taken over that two-dimensional subspace. Its singular values give the principal $g$-factor magnitudes.

The solutions with $g_{\perp,1} \sim 1$ are not rare within these constraints. The representative solution used in the main text includes $B_6^4=0.072\,\mu\mathrm{eV}$ and $B_6^6=0.088\,\mu\mathrm{eV}$ in addition to the fixed rank-2 coefficients. It yields $g_{\perp,1}\approx0.9$, $g_{\perp,2}\approx0.3$, and $g_z\approx16.1$, with a saturation moment $g_z\mu_B/2\approx8.1\,\mu_B$ and a CEF gap of approximately $9.1\,\mathrm{meV}$. This construction demonstrates that appreciable transverse coupling is compatible with the measured constraints; it does not uniquely determine the CEF Hamiltonian or the actual transverse response of ErB$_4$.

\subsubsection*{Density matrix renormalization group}
The DMRG~\cite{White1992, White1993} calculations were performed using the ITensor package \cite{itensor}. For the phase diagrams in Fig.~\ref{fig:Fig1_DMRG}, the geometry used was a long $N_x= 14$, $N_y=6$ cylinder (with open boundary conditions along the shorter edge), with Rydberg interaction tails beyond $V_4$ being dropped as irrelevant compared to $\Omega$. The phase boundaries were reported based on the ridges of entanglement-entropy, with results matching those from structure-factor order parameters. The phase diagrams for varying $N_x\in [11, 19]$ and $N_y \in [6, 8, 10]$ were also evaluated to assess phase stability and commensurability, as discussed in the Supplementary Information. The sweep protocol employed for DMRG was based on typically $\mathcal{O}(100)$ sweeps with bond dimensions in the range $200$--$2500$, depending on $N_y$. Many sweeps were performed at a small bond dimension ($<100$), with progressively increasing bond dimensions in later stages. The noise term of decreasing magnitude was added for bond dimension ($<100$) to aid convergence. The algorithm was considered converged when discarded weights fell below $10^{-11}$, and the change in the ground state energy after the final bond dimension increase was below $10^{-6}$.

\subsubsection*{Quantum phase Ans\" atze}

Here, we aim to capture the effect of quantum fluctuations in the minimal product state or two-site entangled models. This model fails to recover the quantitative data or the correct quantum criticalities, but it can capture the existence of new phases stabilized by quantum fluctuations and the qualitative shape of the phase diagram. The general product state Ansatz for a given state in its unit cell ($C$) is given by:
\begin{equation}\label{eq:productstateA}
    |\Psi_A [\{\theta_i\}]\rangle= \otimes_{i \in C} \left(\cos{\theta_i} |r_i\rangle + \sin{\theta_i} |g_i\rangle \right),
\end{equation}
 where $\theta_i$ are the only free parameters, using the reality of the Rydberg Hamiltonian to fix  the phase choice.

\begin{figure*}[htb]
  \centering
  \includegraphics[width=0.9\linewidth]{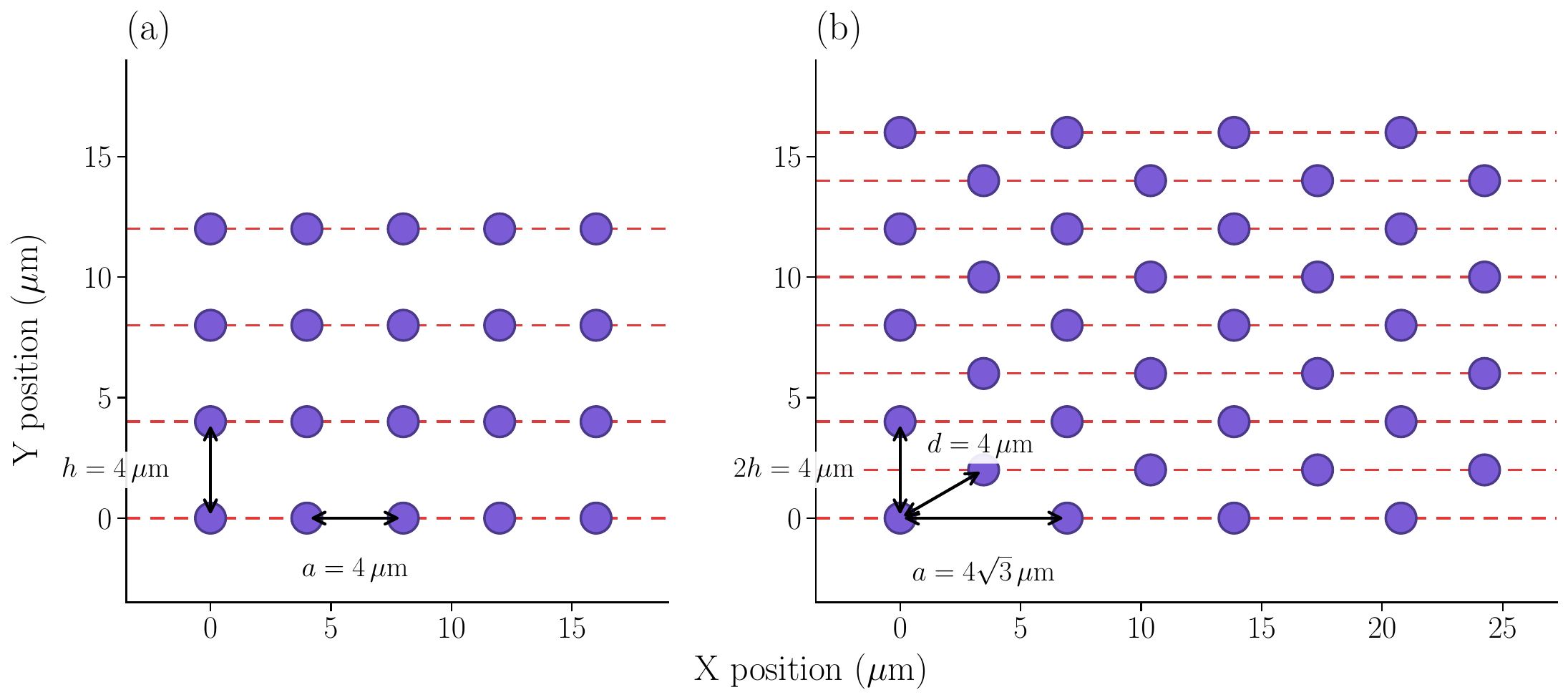}
  \caption{Reservoir packing under the $4\,\mu$m optical-resolution limit.
  (a)~Default square packing: rows and sites are both spaced by $4\,\mu$m.
  (b)~Hexagonal packing at the configuration used here with alternate
  rows offset by $a/2$. The adjacent-row distance $d$ and the skip-row
  vertical distance $2h$ both sit at $4\,\mu$m, giving $a=4\sqrt{3}\,\mu$m and
  $h=2\,\mu$m.}
  \label{fig:reservoir}
\end{figure*}

We then proceed with applying the symmetries of the relevant phases (disordered, 1/3, 2/5, N\' eel, and collinear) in order to fix the number of independent $\theta_i$ within the unit cell. While this fixes the Ansatz for polarized, N\' eel, and collinear phases to the one shown in Fig.~\ref{fig:Fig2_Ansatze}(a), it gives 6 and 10 free parameters for the 1/3 and 2/5 phases, respectively, making the optimization difficult. Taking into account the density profiles obtained in DMRG, we halved  the number of free parameters in these phases by repeating the Ansatz profile in the second part of the unit cell. This slightly disfavors the 1/3 and 2/5 phases over the full-product-state Ansatz, but keeps our qualitative conclusions unchanged.

To capture the stabilization of the quantum dimer regime in a simplified model, one must go beyond the product-state Ansatz, given the phase's essentially entangled nature. We construct a one-parameter Ansatz for the quantum dimer regime motivated by the pure quantum dimer state. In order to include stabilization effects of quantum fluctuations, one needs to admix monomer and double-dimer configurations to the dimer state, leading to:
\begin{align}\label{eq:dimer_Ansatz}
    \ket{\Psi_{A}[\theta]}=&\otimes_{\substack{\mathrm{dimer} \\\mathrm{ bond}\, i}} \frac{1}{\sqrt{2}}\Big[\cos\theta \left(\ket{g_i} \ket{r_{\langle i \rangle^2}} +\ket{r_i} \ket{g_{\langle i \rangle^2}}\right)\\&-\sin\theta \left(\ket{g_i} \ket{g_{\langle i \rangle^2}} +\ket{r_i} \ket{r_{\langle i \rangle^2}}\right)\Big],\notag
\end{align}
where $\theta$ is the mixing angle to be minimized over. Although treating the dimer phase with an entangled two-site Ansatz is not on equal footing with product state Ans\" atze for other phases, it provides a simple mechanism of the dimer phase stabilization in the upper middle region of the phase diagram. This supplement to the DMRG results points to the quantum-dimer character of the unordered middle region.

For each of the Ans\" atze we evaluate the variational energy functional $E[\{\theta_i\}, \Omega, V_k, \Delta]= \bra{\Psi_{A}[\theta]} H_{\mathrm{Ryd}}\ket{\Psi_{A}[\theta]}$. We then use simulated annealing~\cite{Kirkpatrick1983} to determine the lowest-energy state for each point in the phase diagram parameter space. The classical limit corresponds to $\Omega=0$, and it matches the results of Ref.~\cite{Dublenych2013}, as denoted by red dashed lines in Fig.~\ref{fig:Fig1_DMRG}(b). The quantum phase diagrams for strong quantum fluctuation strengths are reported in Fig.~\ref{fig:Fig2_Ansatze}(b)-(c).

\subsection*{Experimental methods}

\subsubsection*{Tall and tight geometries}

On Aquila, atoms are held in static spatial light modulator (SLM) traps and rearranged into the user-specified pattern by acousto-optic deflector (AOD) tweezers that draw atoms from the reservoir regions on the sides of the array~\cite{aquila:2023}. To accelerate the sorting process, atoms are laid out in discrete rows which are rearranged into the final pattern. The optical resolution of the SLM imposes a pairwise radial-distance constraint between any two sites such that no two atoms can be placed within $4\,\mu\mathrm{m}$ of each other~\cite{aquila:2023}. The row-based reservoir packing enforces a minimum \emph{vertical} separation between rows of $4\,\mu\mathrm{m}$~\cite{aquila:2023}. While the minimal radial $4\,\mu\mathrm{m}$ separation is limited by optical resolution, the $4\,\mu\mathrm{m}$ row spacing is a consequence of the reservoir packing strategy rather than an optical limit. 

The total number of traps placed by the SLM is limited by the available laser power. For simplicity, this budget was distributed evenly across all rows independent of their occupancy. As a result, sparsely populated rows received the same number of reservoirs as dense rows which can lead to dense rows becoming reservoir-limited while sparse rows are oversupplied. This effect is rare for short arrays but for taller arrays the per-row budget decreases proportionally and dense rows become the limiting factor on filling fidelity. 

To realize tighter geometries, we replace the default square packing of the reservoir sites (Fig.~\ref{fig:reservoir}a) with a hexagonal (triangular-lattice) packing (Fig.~\ref{fig:reservoir}b). Rather than stacking rows directly above each other, we offset alternate rows horizontally by $a/2$, half the in-row spacing, so that each atom in a given row sits midway between two atoms of the adjacent row.
This converts the inter-row spacing from a vertical constraint into a radial one and the distance between atoms in adjacent rows becomes:
\begin{equation}
    d = \sqrt{\left(\frac{a}{2}\right)^2 + h^2} \geq 4\,\mu\mathrm{m},
    \label{eq:adjacent-row}
\end{equation}
where $h$ is the vertical separation between rows. At the minimum in-row spacing $a = 4\,\mu\mathrm{m}$, $h = \tfrac{\sqrt{3}}{2}\,a = 2\sqrt{3}\,\mu\mathrm{m} \approx 3.46\,\mu\mathrm{m}$. $h$ can be reduced further by widening the in-row spacing $a$. However, with a uniform $a/2$ offset between neighboring rows, atoms separated by two rows share the same horizontal position and are separated only by a vertical distance $2h$. With $2h= 4\,\mu\mathrm{m}$ to respect the optical limit, $a = 4\sqrt{3}\,\mu\mathrm{m} \approx 6.93\,\mu\mathrm{m}$ enabling the tighter arrangements used here at the cost of a wider horizontal spacing (Fig.~\ref{fig:reservoir}b).

To support taller geometries, we replace the uniform allocation with a row-specific allocation derived from the row occupancy and desired filling fidelity. For a row containing $k$ user sites, we solve a binomial reliability problem and find the minimum number of additional reservoir sites $r$ such that the probability of successfully filling all $k$ targets is at least the desired filling fidelity given the per-site loading fidelity. The per-row targets are then scaled by the total available trap budget subject to the maximum number of sites that can physically fit in each row. In the case of deficit, subsequent passes relax the triangular spacing where neighboring rows permit to allow for additional sites. This scheme frees enough sites to support the taller geometries in this work under the same total number of traps.

\subsubsection*{State preparation protocols}

\begin{figure}[htb]
\centering
\includegraphics[width=\linewidth]{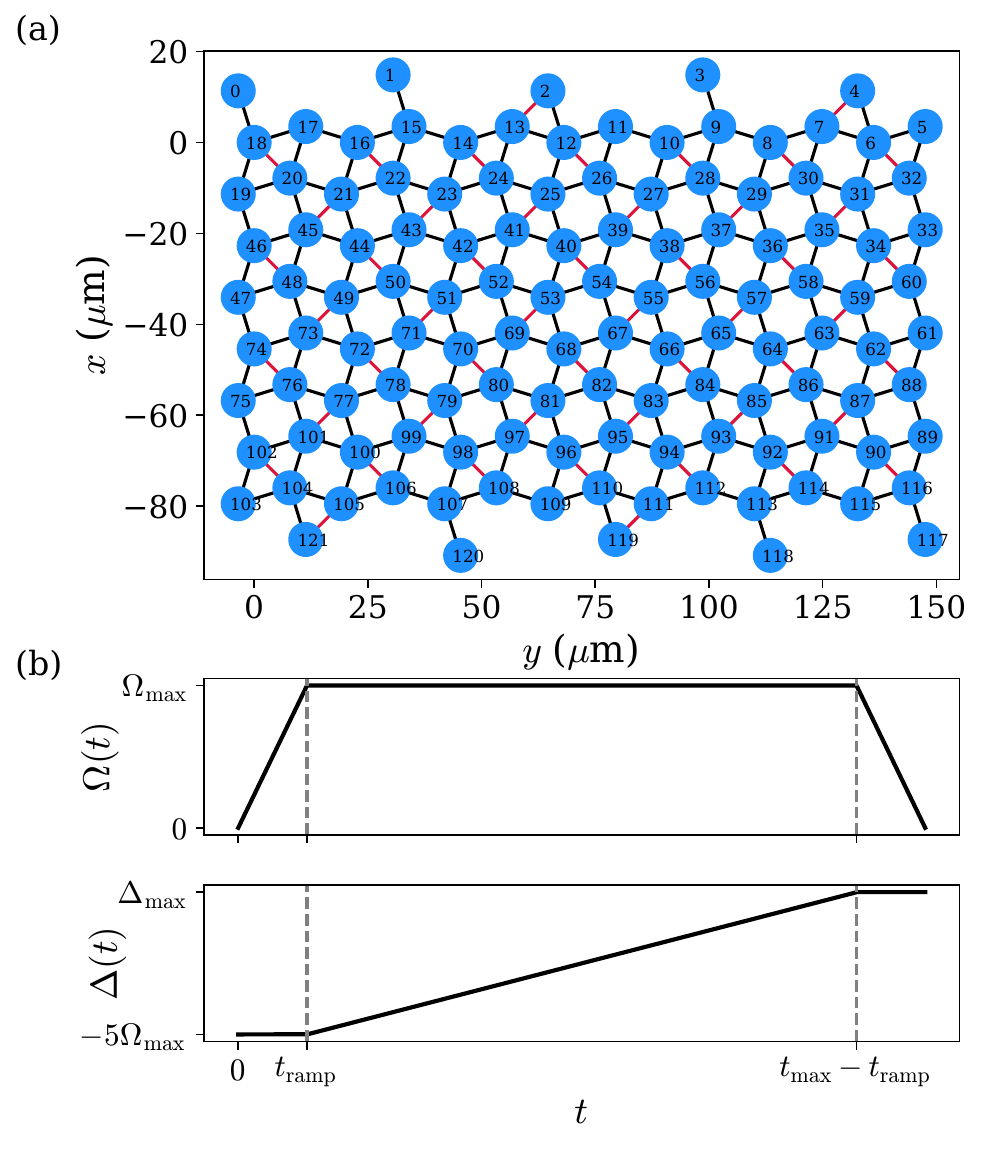}
\caption{Experimental geometry and control protocol.
(a) Example Shastry-Sutherland lattice geometry realized on the Aquila device for a fixed interaction ratio ($V_2/V_1=1.5$ and $R_{b1}=1.1$). (b) Global Rydberg drive protocol used throughout the phase-map measurements. The Rabi frequency ($\Omega(t)$) is smoothly ramped from zero to ($\Omega_{\max}$), held constant during the central evolution period, and subsequently ramped back to zero. Simultaneously, the detuning $\Delta(t)$ is swept from $\Delta_{\min}=-5\Omega_{\max}$ to a final value $\Delta_{\max}$, corresponding to the target point in parameter space $(V_2/V_1,\Delta/V_1)$. Vertical dashed lines indicate the ramp up $t_{\rm ramp}$ and ramp down $t_{\max}-t_{\rm ramp}$ intervals.}
\label{fig:Fig_exp_waveforms}
\end{figure}

All phase-map studies are performed at fixed interaction strength ratio $V_{2}/V_{1}$ while sweeping the detuning $\Delta/V_{1}$ through a time-dependent detuning waveform $\Delta(t)$. The lattice geometry is generated from a parameterized structure fully determined by $V_{2}/V_{1}$, which fixes the Shastry-Sutherland lattice angle, $\alpha = 2 \arcsin\left(\left(V_{1}/V_{2}\right)^{1/6}/2\right)$.
The lattice spacing $R_1$, which sets the interaction scale via $V_{1} = C_6 / R_1^6$, is chosen to satisfy the device constraint on the minimum interatomic separation, $D_{\min} = 4~\mu\mathrm{m}$. This constraint, together with the prescribed dimensionless blockade radius $R_{b1}$, determines the admissible Rabi frequency $\Omega$. Each fixed value of $V_{2}/V_{1}$ therefore defines a unique and fully specified lattice geometry, an example of which is shown on Fig.~\ref{fig:Fig_exp_waveforms}a.  To reconstruct the full phase diagram, this procedure is repeated for multiple values of $V_{2}/V_{1}$, thereby sampling the desired parameter space.

In experiments, the control fields are implemented as continuous  piecewise-linear pulse sequences as shown in Fig.~\ref{fig:Fig_exp_waveforms}b~\cite{bloqadepython:2024, BloqadeJulia:2023}. The Rabi frequency $\Omega(t)$ is shaped as a finite-duration ramp protocol in which the system is initially prepared in a non-driven state, ramped up to a maximum value $\Omega$, held constant during the bulk evolution, and finally ramped down symmetrically to zero. Specifically, $\Omega(t)$ increases linearly from $0$ to $\Omega_{\rm max}$. over a ramp time $t_{\rm ramp}$, remains constant for the central portion of the sequence, and is then decreased linearly back to zero over the same duration. The plateau value is set by the blockade ratio through $\Omega = C_6/(R_{b1}R_{1})^6$. For the boundary-pinned phase diagrams at $R_{b1}=1.1$ it ranges over $\Omega/2\pi = 1.43$--$1.83$~MHz across $V_2/V_1 = 0.9$--$2.0$ in all cases below the hardware maximum $\Omega_{\rm max}/2\pi = 2.5$~MHz. The finite turn-on and turn-off ramps avoid abrupt control discontinuities while attempting to minimize non-adiabatic effects.

The detuning $\Delta(t)$ is held at an initial value $\Delta_{\min} = -5\Omega_{\max}$ during the initial ramp, then increased linearly during the central evolution window to a final value $\Delta_{\max}$ corresponding to the target $\Delta/V_{1}$, and finally held constant for the remainder of the sequence. The total protocol duration is set to the maximum evolution time allowed by the hardware at $t_{\max} = 4~\mu\mathrm{s}$. The detuning and Rabi waveforms are discretized with a minimum time resolution $\Delta t_{\min} = 50~\mathrm{ns}$. Accordingly, the ramp duration is chosen as $t_{\mathrm{ramp}} = 5\Delta t_{\min} = 250~\mathrm{ns}$, and  ensuring both smooth boundary ramps and maximal utilization of the evolution window.

\subsubsection*{Data analysis}

Following the state-preparation sweep, the atomic array is measured in the Rydberg occupation basis over many experimental repetitions.  For each parameter pair $(V_{2}/V_{1},\Delta/V_{1})$, 500 experimental shots are acquired. Each experimental shot yields a binary configuration $\{n_i\}$, where $n_i=1$ ($0$) denotes the presence (absence) of a Rydberg excitation on site $i$. All observables presented in this work are extracted directly from these measured configurations. 
Site-resolved densities are obtained from the ensemble averages $\langle n_i\rangle$, while bond and dimer observables are constructed from occupations on neighboring lattice sites. For each point in the parameter space $(V_2/V_1,\Delta/V_1)$, dimer order parameters and bond-resolved correlation functions are evaluated for every experimental realization and subsequently averaged over the full measurement ensemble.
To characterize long-range ordering, static structure factors are computed from the measured Rydberg density correlations by Fourier transforming the corresponding real-space occupancy correlation functions according to Eq.~\ref{eq:SSF}. The amplitudes of the resulting structure factors at the dominant ordering wavevectors are used as order parameters to construct the phase maps and identify the different ordered phases across the phase diagram. Statistical uncertainties are estimated from shot-to-shot fluctuations and propagated to all derived observables.

\bibliography{SSL.bib}

@misc{AWSBraket2020,
  author = {{Amazon Web Services}},
  title  = {{Amazon Braket}},
  year   = {2020},
  url    = {https://aws.amazon.com/braket/},   
}

@article{Korzhenevskii1975,
  title = {Connection between the properties of the system of renormalization-group equations and the symmetry of the Hamiltonian in the phase-transition problem},
  author = {Korzhenevskii, A.L.},
  journal = {JETP},
  volume = {44},
  issue = {4},
  pages = {751},
  year = {1976},
  month = {Nov},
  url = {http://jetp.ras.ru/cgi-bin/dn/e_044_04_0751.pdf}
}

@article{Kalinowski2022,
  title = {Bulk and boundary quantum phase transitions in a square Rydberg atom array},
  author = {Kalinowski, Marcin and Samajdar, Rhine and Melko, Roger G. and Lukin, Mikhail D. and Sachdev, Subir and Choi, Soonwon},
  journal = {Phys. Rev. B},
  volume = {105},
  issue = {17},
  pages = {174417},
  numpages = {15},
  year = {2022},
  month = {May},
  publisher = {American Physical Society},
  doi = {10.1103/PhysRevB.105.174417},
  url = {https://link.aps.org/doi/10.1103/PhysRevB.105.174417}
}

@article{Aharony1973,
  title = {Critical Behavior of Anisotropic Cubic Systems},
  author = {Aharony, Amnon},
  journal = {Phys. Rev. B},
  volume = {8},
  issue = {9},
  pages = {4270--4273},
  numpages = {0},
  year = {1973},
  month = {Nov},
  publisher = {American Physical Society},
  doi = {10.1103/PhysRevB.8.4270},
  url = {https://link.aps.org/doi/10.1103/PhysRevB.8.4270}
}

@article{Grinstein1976,
  title = {Application of the renormalization group to phase transitions in disordered systems},
  author = {Grinstein, G. and Luther, A.},
  journal = {Phys. Rev. B},
  volume = {13},
  issue = {3},
  pages = {1329--1343},
  numpages = {0},
  year = {1976},
  month = {Feb},
  publisher = {American Physical Society},
  doi = {10.1103/PhysRevB.13.1329},
  url = {https://link.aps.org/doi/10.1103/PhysRevB.13.1329}
}

@article{Kawamura1988,
  title = {Renormalization-group analysis of chiral transitions},
  author = {Kawamura, Hikaru},
  journal = {Phys. Rev. B},
  volume = {38},
  issue = {7},
  pages = {4916--4928},
  numpages = {0},
  year = {1988},
  month = {Sep},
  publisher = {American Physical Society},
  doi = {10.1103/PhysRevB.38.4916},
  url = {https://link.aps.org/doi/10.1103/PhysRevB.38.4916}
}

@article{Janke1997,
title = {Three-dimensional 3-state Potts model revisited with new techniques},
journal = {Nuclear Physics B},
volume = {489},
number = {3},
pages = {679-696},
year = {1997},
issn = {0550-3213},
doi = {https://doi.org/10.1016/S0550-3213(96)00710-9},
url = {https://www.sciencedirect.com/science/article/pii/S0550321396007109},
author = {Wolfhard Janke and Ramon Villanova}
}

@Article{El-Showk2014,
author={El-Showk, Sheer
and Paulos, Miguel F.
and Poland, David
and Rychkov, Slava
and Simmons-Duffin, David
and Vichi, Alessandro},
title={Solving the 3d Ising Model with the Conformal Bootstrap II. {\$}{\$}c{\$}{\$}-Minimization and Precise Critical Exponents},
journal={Journal of Statistical Physics},
year={2014},
month={Dec},
day={01},
volume={157},
number={4},
pages={869-914},
issn={1572-9613},
doi={10.1007/s10955-014-1042-7},
url={https://doi.org/10.1007/s10955-014-1042-7}
}

@INPROCEEDINGS{Jha2021,
  author={Jha, Akshat A. and Stoyanoff, Eliana L. and Khundzakishvili, Guga and Kairys, Paul and Ushijima-Mwesigwa, Hayato and Banerjee, Arnab},
  booktitle={2021 International Conference on Rebooting Computing (ICRC)}, 
  title={Digital Annealing Route to Complex Magnetic Phase Discovery}, 
  year={2021},
  volume={},
  number={},
  pages={119-123},
  doi={10.1109/ICRC53822.2021.00027}}

@article{Pelissetto2002,
title = {Critical phenomena and renormalization-group theory},
journal = {Physics Reports},
volume = {368},
number = {6},
pages = {549-727},
year = {2002},
issn = {0370-1573},
doi = {https://doi.org/10.1016/S0370-1573(02)00219-3},
url = {https://www.sciencedirect.com/science/article/pii/S0370157302002193},
author = {Andrea Pelissetto and Ettore Vicari}
}

@misc{itensor,
  title={The \mbox{ITensor} Software Library for Tensor Network Calculations},
  author={Matthew Fishman and Steven R. White and E. Miles Stoudenmire},
  year={2020},
  eprint={2007.14822},
  archivePrefix={arXiv}
}

@article{White1993,
  title = {Density-matrix algorithms for quantum renormalization groups},
  author = {White, Steven R.},
  journal = {Phys. Rev. B},
  volume = {48},
  issue = {14},
  pages = {10345--10356},
  numpages = {0},
  year = {1993},
  month = {Oct},
  publisher = {American Physical Society},
  doi = {10.1103/PhysRevB.48.10345},
  url = {https://link.aps.org/doi/10.1103/PhysRevB.48.10345}
}

@article{Li2022,
  title    = "Quantum phases of Rydberg atoms on a frustrated
              triangular-lattice array",
  author   = "Li, Chang-Xiao and Yang, Sheng and Xu, Jing-Bo",
  journal  = "Opt Lett",
  volume   =  47,
  number   =  5,
  pages    = "1093--1096",
  month    =  mar,
  year     =  2022,
  address  = "United States"
  }

@article{Samajdar2021,
author = {Rhine Samajdar  and Wen Wei Ho  and Hannes Pichler  and Mikhail D. Lukin  and Subir Sachdev },
title = {Quantum phases of Rydberg atoms on a kagome lattice},
journal = {Proceedings of the National Academy of Sciences},
volume = {118},
number = {4},
pages = {e2015785118},
year = {2021},
doi = {10.1073/pnas.2015785118},
URL = {https://www.pnas.org/doi/abs/10.1073/pnas.2015785118},
eprint = {https://www.pnas.org/doi/pdf/10.1073/pnas.2015785118}}

@misc{bloqadepython:2024,
  author       = {Weinberg, Phillip and Wu, Kai-Hsin and John Long and Luo, Xiu-zhe (Roger)},
  title        = {QuEraComputing/bloqade-python: v0.15.11},
  month        = may,
  year         = 2024,
  publisher    = {Zenodo},
  version      = {v0.15.11},
  doi          = {10.5281/zenodo.11114110},
  url          = {https://doi.org/10.5281/zenodo.11114110}
}

@misc{BloqadeJulia:2023,
  url = {https://github.com/QuEraComputing/Bloqade.jl/},
  title = {Bloqade.jl: {P}ackage for the quantum computation and quantum simulation based on the neutral-atom architecture.},
  year = {2023}
}

@BOOK{Fazekas1999,
  title     = "Lecture notes on electron correlation and magnetism",
  author    = "Fazekas, Patrick",
  publisher = "World Scientific Publishing Co. Pte. Ltd.",
  year      =  1999
}

@article{Kirkpatrick1983,
author = {S. Kirkpatrick  and C. D. Gelatt  and M. P. Vecchi },
title = {Optimization by Simulated Annealing},
journal = {Science},
volume = {220},
number = {4598},
pages = {671-680},
year = {1983},
doi = {10.1126/science.220.4598.671},
URL = {https://www.science.org/doi/abs/10.1126/science.220.4598.671},
eprint = {https://www.science.org/doi/pdf/10.1126/science.220.4598.671}}

@article{Rourke2023,
  title="{Entanglement in the quantum phases of an unfrustrated Rydberg atom array}",
  author={O’Rourke, Matthew J and Chan, Garnet Kin-Lic},
  journal={Nat. Commun.},
  volume={14},
  number={1},
  pages={5397},
  year={2023},
  publisher={Nature Publishing Group UK London},
  doi={10.1038/s41467-023-41166-0}
}

@article{White1992,
  title = {Density matrix formulation for quantum renormalization groups},
  author = {White, Steven R.},
  journal = {Phys. Rev. Lett.},
  volume = {69},
  issue = {19},
  pages = {2863--2866},
  numpages = {0},
  year = {1992},
  month = {Nov},
  publisher = {American Physical Society},
  doi = {10.1103/PhysRevLett.69.2863},
  url = {https://link.aps.org/doi/10.1103/PhysRevLett.69.2863}
}

@article{Stevens1952,
doi = {10.1088/0370-1298/65/3/308},
url = {https://doi.org/10.1088/0370-1298/65/3/308},
year = {1952},
month = {mar},
publisher = {},
volume = {65},
number = {3},
pages = {209},
author = {K W H Stevens},
title = {Matrix Elements and Operator Equivalents Connected with the Magnetic Properties of Rare Earth Ions},
journal = {Proceedings of the Physical Society. Section A}
}

@Article{Labuhn2016,
author={Labuhn, Henning
and Barredo, Daniel
and Ravets, Sylvain
and de L{\'e}s{\'e}leuc, Sylvain
and Macr{\`i}, Tommaso
and Lahaye, Thierry
and Browaeys, Antoine},
title={Tunable two-dimensional arrays of single Rydberg atoms for realizing quantum Ising models},
journal={Nature},
year={2016},
month={Jun},
day={01},
volume={534},
number={7609},
pages={667-670},
issn={1476-4687},
doi={10.1038/nature18274},
url={https://doi.org/10.1038/nature18274}
}

@article{Semeghini2021,
author = {G. Semeghini  and H. Levine  and A. Keesling  and S. Ebadi  and T. T. Wang  and D. Bluvstein  and R. Verresen  and H. Pichler  and M. Kalinowski  and R. Samajdar  and A. Omran  and S. Sachdev  and A. Vishwanath  and M. Greiner  and V. Vuletić  and M. D. Lukin },
title = {Probing topological spin liquids on a programmable quantum simulator},
journal = {Science},
volume = {374},
number = {6572},
pages = {1242-1247},
year = {2021},
doi = {10.1126/science.abi8794},
URL = {https://www.science.org/doi/abs/10.1126/science.abi8794},
eprint = {https://www.science.org/doi/pdf/10.1126/science.abi8794}}

@Article{Ebadi2021,
author={Ebadi, Sepehr
and Wang, Tout T.
and Levine, Harry
and Keesling, Alexander
and Semeghini, Giulia
and Omran, Ahmed
and Bluvstein, Dolev
and Samajdar, Rhine
and Pichler, Hannes
and Ho, Wen Wei
and Choi, Soonwon
and Sachdev, Subir
and Greiner, Markus
and Vuleti{\'{c}}, Vladan
and Lukin, Mikhail D.},
title={Quantum phases of matter on a 256-atom programmable quantum simulator},
journal={Nature},
year={2021},
month={Jul},
day={01},
volume={595},
number={7866},
pages={227-232},
issn={1476-4687},
doi={10.1038/s41586-021-03582-4},
url={https://doi.org/10.1038/s41586-021-03582-4}
}

@article{Shastry1981,
title = {Exact ground state of a quantum mechanical antiferromagnet},
journal = {Physica B+C},
volume = {108},
number = {1},
pages = {1069-1070},
year = {1981},
issn = {0378-4363},
doi = {https://doi.org/10.1016/0378-4363(81)90838-X},
url = {https://www.sciencedirect.com/science/article/pii/037843638190838X},
author = {B. {Sriram Shastry} and Bill Sutherland}
}

@article{Kairys2020,  
  title = {Simulating the Shastry-Sutherland Ising Model Using Quantum Annealing},
  author = {Kairys, Paul and King, Andrew D. and Ozfidan, Isil and Boothby, Kelly and Raymond, Jack and Banerjee, Arnab and Humble, Travis S.},
  journal = {PRX Quantum},
  volume = {1},
  issue = {2},
  pages = {020320},
  numpages = {11},
  year = {2020},
  month = {Dec},
  publisher = {American Physical Society},
  doi = {10.1103/PRXQuantum.1.020320},
  url = {https://link.aps.org/doi/10.1103/PRXQuantum.1.020320}
}

@article{Dublenych2013,
  title = {Ground states of an Ising model on an extended Shastry-Sutherland lattice and the $1/2$-magnetization plateau in some rare-earth-metal tetraborides},
  author = {Dublenych, Yu. I.},
  journal = {Phys. Rev. E},
  volume = {88},
  issue = {2},
  pages = {022111},
  numpages = {13},
  year = {2013},
  month = {Aug},
  publisher = {American Physical Society},
  doi = {10.1103/PhysRevE.88.022111},
  url = {https://link.aps.org/doi/10.1103/PhysRevE.88.022111}
}

@article{Meng2008,
  title = {Phases and magnetization process of an anisotropic Shastry-Sutherland model},
  author = {Meng, Zi Yang and Wessel, Stefan},
  journal = {Phys. Rev. B},
  volume = {78},
  issue = {22},
  pages = {224416},
  numpages = {12},
  year = {2008},
  month = {Dec},
  publisher = {American Physical Society},
  doi = {10.1103/PhysRevB.78.224416},
  url = {https://link.aps.org/doi/10.1103/PhysRevB.78.224416}
}

@article{Dublenych2012,
  title = {Ground States of the Ising Model on the Shastry-Sutherland Lattice and the Origin of the Fractional Magnetization Plateaus in Rare-Earth-Metal Tetraborides},
  author = {Dublenych, Yu. I.},
  journal = {Phys. Rev. Lett.},
  volume = {109},
  issue = {16},
  pages = {167202},
  numpages = {5},
  year = {2012},
  month = {Oct},
  publisher = {American Physical Society},
  doi = {10.1103/PhysRevLett.109.167202},
  url = {https://link.aps.org/doi/10.1103/PhysRevLett.109.167202}
}

@article{Dublenych2014,
  title = {Ground-state structures in Ising magnets on the Shastry-Sutherland lattice with long-range interactions and fractional magnetization plateaus in ${\mathrm{TmB}}_{4}$},
  author = {Dublenych, Yu. I.},
  journal = {Phys. Rev. E},
  volume = {90},
  issue = {5},
  pages = {052123},
  numpages = {18},
  year = {2014},
  month = {Nov},
  publisher = {American Physical Society},
  doi = {10.1103/PhysRevE.90.052123},
  url = {https://link.aps.org/doi/10.1103/PhysRevE.90.052123}
}

@article{Yang2022,
  title = {Quantum criticality and spin liquid phase in the Shastry-Sutherland model},
  author = {Yang, Jianwei and Sandvik, Anders W. and Wang, Ling},
  journal = {Phys. Rev. B},
  volume = {105},
  issue = {6},
  pages = {L060409},
  numpages = {7},
  year = {2022},
  month = {Feb},
  publisher = {American Physical Society},
  doi = {10.1103/PhysRevB.105.L060409},
  url = {https://link.aps.org/doi/10.1103/PhysRevB.105.L060409}
}

@article{Lee2019,
  title = {Signatures of a Deconfined Phase Transition on the Shastry-Sutherland Lattice: Applications to Quantum Critical ${\mathrm{SrCu}}_{2}({\mathrm{BO}}_{3}{)}_{2}$},
  author = {Lee, Jong Yeon and You, Yi-Zhuang and Sachdev, Subir and Vishwanath, Ashvin},
  journal = {Phys. Rev. X},
  volume = {9},
  issue = {4},
  pages = {041037},
  numpages = {25},
  year = {2019},
  month = {Nov},
  publisher = {American Physical Society},
  doi = {10.1103/PhysRevX.9.041037},
  url = {https://link.aps.org/doi/10.1103/PhysRevX.9.041037}
}

@article{Kageyama1999,
  title = {Exact Dimer Ground State and Quantized Magnetization Plateaus in the Two-Dimensional Spin System ${\mathrm{SrCu}}_{2}({\mathrm{BO}}_{3}){}_{2}$},
  author = {Kageyama, H. and Yoshimura, K. and Stern, R. and Mushnikov, N. V. and Onizuka, K. and Kato, M. and Kosuge, K. and Slichter, C. P. and Goto, T. and Ueda, Y.},
  journal = {Phys. Rev. Lett.},
  volume = {82},
  issue = {15},
  pages = {3168--3171},
  numpages = {0},
  year = {1999},
  month = {Apr},
  publisher = {American Physical Society},
  doi = {10.1103/PhysRevLett.82.3168},
  url = {https://link.aps.org/doi/10.1103/PhysRevLett.82.3168}
}

@article{Siemensmeyer2008,
  title = {Fractional Magnetization Plateaus and Magnetic Order in the Shastry-Sutherland Magnet ${\mathrm{TmB}}_{4}$},
  author = {Siemensmeyer, K. and Wulf, E. and Mikeska, H.-J. and Flachbart, K. and Gab\'ani, S. and Mat'a\ifmmode \check{s}\else \v{s}\fi{}, S. and Priputen, P. and Efdokimova, A. and Shitsevalova, N.},
  journal = {Phys. Rev. Lett.},
  volume = {101},
  issue = {17},
  pages = {177201},
  numpages = {4},
  year = {2008},
  month = {Oct},
  publisher = {American Physical Society},
  doi = {10.1103/PhysRevLett.101.177201},
  url = {https://link.aps.org/doi/10.1103/PhysRevLett.101.177201}
}

@article{Yoshii_2006,
doi = {10.1088/1742-6596/51/1/011},
url = {https://doi.org/10.1088/1742-6596/51/1/011},
year = {2006},
month = {nov},
publisher = {},
volume = {51},
number = {1},
pages = {59},
author = {S Yoshii and T Yamamoto and M Hagiwara and A Shigekawa and S Michimura and F Iga and T Takabatake and K Kindo},
title = {High-field magnetization of TmB4},
journal = {Journal of Physics: Conference Series}
}

@Article{Keesling2019,
author={Keesling, Alexander
and Omran, Ahmed
and Levine, Harry
and Bernien, Hannes
and Pichler, Hannes
and Choi, Soonwon
and Samajdar, Rhine
and Schwartz, Sylvain
and Silvi, Pietro
and Sachdev, Subir
and Zoller, Peter
and Endres, Manuel
and Greiner, Markus
and Vuleti{\'{c}}, Vladan
and Lukin, Mikhail D.},
title={Quantum Kibble--Zurek mechanism and critical dynamics on a programmable Rydberg simulator},
journal={Nature},
year={2019},
month={Apr},
day={01},
volume={568},
number={7751},
pages={207-211},
issn={1476-4687},
doi={10.1038/s41586-019-1070-1},
url={https://doi.org/10.1038/s41586-019-1070-1}
}

@Article{Balents2010,
author={Balents, Leon},
title={Spin liquids in frustrated magnets},
journal={Nature},
year={2010},
month={Mar},
day={01},
volume={464},
number={7286},
pages={199-208},
issn={1476-4687},
doi={10.1038/nature08917},
url={https://doi.org/10.1038/nature08917}
}

@BOOK{Lacroix2011,
  title     = "Introduction to frustrated magnetism",
  editor    = "Lacroix, Claudine and Mila, Frederic and Mendels, Philippe",
  publisher = "Springer",
  series    = "Springer Series in Solid-State Sciences",
  edition   =  2011,
  month     =  {jan},
  year      =  2011
}

@article{Chubukov1991,
doi = {10.1088/0953-8984/3/1/005},
url = {https://doi.org/10.1088/0953-8984/3/1/005},
year = {1991},
month = {jan},
publisher = {},
volume = {3},
number = {1},
pages = {69},
author = {A V Chubukov and D I Golosov},
title = {Quantum theory of an antiferromagnet on a triangular lattice in a magnetic field},
journal = {Journal of Physics: Condensed Matter}
}

@misc{geim2026engineeringquantumcriticalitydynamics,
      title={Engineering quantum criticality and dynamics on an analog-digital simulator}, 
      author={Alexandra A. Geim and Nazli Ugur Koyluoglu and Simon J. Evered and Rahul Sahay and Sophie H. Li and Muqing Xu and Dolev Bluvstein and Nik O. Gjonbalaj and Nishad Maskara and Marcin Kalinowski and Tom Manovitz and Ruben Verresen and Susanne F. Yelin and Johannes Feldmeier and Markus Greiner and Vladan Vuletic and Mikhail D. Lukin},
      year={2026},
      eprint={2602.18555},
      archivePrefix={arXiv},
      primaryClass={quant-ph},
      url={https://arxiv.org/abs/2602.18555}, 
}

@article{Alicea2009,
  title = {Quantum Stabilization of the $1/3$-Magnetization Plateau in ${\mathrm{Cs}}_{2}{\mathrm{CuBr}}_{4}$},
  author = {Alicea, Jason and Chubukov, Andrey V. and Starykh, Oleg A.},
  journal = {Phys. Rev. Lett.},
  volume = {102},
  issue = {13},
  pages = {137201},
  numpages = {4},
  year = {2009},
  month = {Mar},
  publisher = {American Physical Society},
  doi = {10.1103/PhysRevLett.102.137201},
  url = {https://link.aps.org/doi/10.1103/PhysRevLett.102.137201}
}

@article{Kawamura1985,
author = {Kawamura ,Hikaru and Miyashita ,Seiji},
title = {Phase Transition of the Heisenberg Antiferromagnet on the Triangular Lattice in a Magnetic Field},
journal = {Journal of the Physical Society of Japan},
volume = {54},
number = {12},
pages = {4530-4538},
year = {1985},
doi = {10.1143/JPSJ.54.4530},
URL = {https://doi.org/10.1143/JPSJ.54.4530},
eprint = { https://doi.org/10.1143/JPSJ.54.4530}
}

@article{Savary2017,
doi = {10.1088/0034-4885/80/1/016502},
url = {https://doi.org/10.1088/0034-4885/80/1/016502},
year = {2016},
month = {nov},
publisher = {IOP Publishing},
volume = {80},
number = {1},
pages = {016502},
author = {Savary, Lucile and Balents, Leon},
title = {Quantum spin liquids: a review},
journal = {Reports on Progress in Physics}
}

@article{Wannier1950,
  title = {Antiferromagnetism. The Triangular Ising Net},
  author = {Wannier, G. H.},
  journal = {Phys. Rev.},
  volume = {79},
  issue = {2},
  pages = {357--364},
  numpages = {0},
  year = {1950},
  month = {Jul},
  publisher = {American Physical Society},
  doi = {10.1103/PhysRev.79.357},
  url = {https://link.aps.org/doi/10.1103/PhysRev.79.357}
}

@article{Moessner2001,
  title = {Ising models of quantum frustration},
  author = {Moessner, R. and Sondhi, S. L.},
  journal = {Phys. Rev. B},
  volume = {63},
  issue = {22},
  pages = {224401},
  numpages = {19},
  year = {2001},
  month = {May},
  publisher = {American Physical Society},
  doi = {10.1103/PhysRevB.63.224401},
  url = {https://link.aps.org/doi/10.1103/PhysRevB.63.224401}
}

@article{Harris1997,
  title = {Geometrical Frustration in the Ferromagnetic Pyrochlore ${\mathrm{Ho}}_{2}{\mathrm{Ti}}_{2}{O}_{7}$},
  author = {Harris, M. J. and Bramwell, S. T. and McMorrow, D. F. and Zeiske, T. and Godfrey, K. W.},
  journal = {Phys. Rev. Lett.},
  volume = {79},
  issue = {13},
  pages = {2554--2557},
  numpages = {0},
  year = {1997},
  month = {Sep},
  publisher = {American Physical Society},
  doi = {10.1103/PhysRevLett.79.2554},
  url = {https://link.aps.org/doi/10.1103/PhysRevLett.79.2554}
}

@article{Trinh2018,
  title = {Degeneracy of the $1/8$ Plateau and Antiferromagnetic Phases in the Shastry-Sutherland Magnet ${\mathrm{TmB}}_{4}$},
  author = {Trinh, Jennifer and Mitra, Sreemanta and Panagopoulos, Christos and Kong, Tai and Canfield, Paul C. and Ramirez, Arthur P.},
  journal = {Phys. Rev. Lett.},
  volume = {121},
  issue = {16},
  pages = {167203},
  numpages = {6},
  year = {2018},
  month = {Oct},
  publisher = {American Physical Society},
  doi = {10.1103/PhysRevLett.121.167203},
  url = {https://link.aps.org/doi/10.1103/PhysRevLett.121.167203}
}

@article{Matas2010,
doi = {10.1088/1742-6596/200/3/032041},
url = {https://doi.org/10.1088/1742-6596/200/3/032041},
year = {2010},
month = {jan},
publisher = {},
volume = {200},
number = {3},
pages = {032041},
author = {S Mat'aš and K Siemensmeyer and E Wheeler and E Wulf and R Beyer and Th Hermannsdörfer and O Ignatchik and M Uhlarz and K Flachbart and S Gabáni and P Priputen and A Efdokimova and N Shitsevalova},
title = {Magnetism of rare earth tetraborides},
journal = {Journal of Physics: Conference Series}
}

@article{Etourneau1985,
author = {J. Etourneau and P. Hagenmuller},
title = {Structure and physical features of the rare-earth borides},
journal = {Philosophical Magazine B},
volume = {52},
number = {3},
pages = {589--610},
year = {1985},
publisher = {Taylor \& Francis},
doi = {10.1080/13642818508240625},
URL = {https://doi.org/10.1080/13642818508240625},
eprint = {https://doi.org/10.1080/13642818508240625}
}

@article{Ye2017,
  title = {Electronic transport on the Shastry-Sutherland lattice in Ising-type rare-earth tetraborides},
  author = {Ye, Linda and Suzuki, Takehito and Checkelsky, Joseph G.},
  journal = {Phys. Rev. B},
  volume = {95},
  issue = {17},
  pages = {174405},
  numpages = {7},
  year = {2017},
  month = {May},
  publisher = {American Physical Society},
  doi = {10.1103/PhysRevB.95.174405},
  url = {https://link.aps.org/doi/10.1103/PhysRevB.95.174405}
}

@article{Flury2025,
  title = {Magnetic phase diagram of ErB4 as explored by neutron scattering},
  author = {Flury, Simon and Simeth, Wolfgang J. and Yahne, Danielle R. and Islam, Manisha and Plokhikh, Igor and Mazzone, Daniel G. and Bauer, Eric D. and Rosa, Priscila F. S. and Sibille, Romain and Zaharko, Oksana and Gawryluk, Dariusz J. and Janoschek, Marc},
  journal = {Phys. Rev. B},
  volume = {112},
  issue = {22},
  pages = {224441},
  numpages = {11},
  year = {2025},
  month = {Dec},
  publisher = {American Physical Society},
  doi = {10.1103/m3wx-4v6k},
  url = {https://link.aps.org/doi/10.1103/m3wx-4v6k}
}

@article{Michimura2006,
title = {Magnetic frustrations in the Shastry–Sutherland system ErB4},
journal = {Physica B: Condensed Matter},
volume = {378-380},
pages = {596-597},
year = {2006},
note = {Proceedings of the International Conference on Strongly Correlated Electron Systems},
issn = {0921-4526},
doi = {https://doi.org/10.1016/j.physb.2006.01.161},
url = {https://www.sciencedirect.com/science/article/pii/S0921452606002857},
author = {S. Michimura and A. Shigekawa and F. Iga and M. Sera and T. Takabatake and K. Ohoyama and Y. Okabe}
}

@article{Schafer1976,
    author = {Schäfer, W. and Will, G. and Buschow, K. H. J.},
    title = {The magnetic structure of the rare earth tetraborides ErB4 and DyB4},
    journal = {The Journal of Chemical Physics},
    volume = {64},
    number = {5},
    pages = {1994-1997},
    year = {1976},
    month = {03},
    issn = {0021-9606},
    doi = {10.1063/1.432464},
    url = {https://doi.org/10.1063/1.432464}
}

@article{Wierschem2013,
  title = {Columnar Antiferromagnetic Order and Spin Supersolid Phase on the Extended Shastry-Sutherland Lattice},
  author = {Wierschem, Keola and Sengupta, Pinaki},
  journal = {Phys. Rev. Lett.},
  volume = {110},
  issue = {20},
  pages = {207207},
  numpages = {5},
  year = {2013},
  month = {May},
  publisher = {American Physical Society},
  doi = {10.1103/PhysRevLett.110.207207},
  url = {https://link.aps.org/doi/10.1103/PhysRevLett.110.207207}
}

@article{Farkas2010,
  title = {Numerical study of magnetization processes in rare-earth tetraborides},
  author = {Farka\v{s}ovsk\'y, Pavol and \v{C}en\v{c}arikov\'a, Hana and Mat'a\v{s}, Slavom\'{\i}r},
  journal = {Phys. Rev. B},
  volume = {82},
  issue = {5},
  pages = {054409},
  numpages = {5},
  year = {2010},
  month = {Aug},
  publisher = {American Physical Society},
  doi = {10.1103/PhysRevB.82.054409},
  url = {https://link.aps.org/doi/10.1103/PhysRevB.82.054409}
}

@article{Bluvstein:2021,
author = {D. Bluvstein  and A. Omran  and H. Levine  and A. Keesling  and G. Semeghini  and S. Ebadi  and T. T. Wang  and A. A. Michailidis  and N. Maskara  and W. W. Ho  and S. Choi  and M. Serbyn  and M. Greiner  and V. Vuletić  and M. D. Lukin },
title = {Controlling quantum many-body dynamics in driven Rydberg atom arrays},
journal = {Science},
volume = {371},
number = {6536},
pages = {1355-1359},
year = {2021},
doi = {10.1126/science.abg2530},
url = {https://www.science.org/doi/abs/10.1126/science.abg2530}
}

@Article{Bernien2017,
author={Bernien, Hannes
and Schwartz, Sylvain
and Keesling, Alexander
and Levine, Harry
and Omran, Ahmed
and Pichler, Hannes
and Choi, Soonwon
and Zibrov, Alexander S.
and Endres, Manuel
and Greiner, Markus
and Vuleti{\'{c}}, Vladan
and Lukin, Mikhail D.},
title={Probing many-body dynamics on a 51-atom quantum simulator},
journal={Nature},
year={2017},
month={Nov},
day={01},
volume={551},
number={7682},
pages={579-584},
issn={1476-4687},
doi={10.1038/nature24622},
url={https://doi.org/10.1038/nature24622}
}

@misc{aquila:2023,
      title={Aquila: QuEra's 256-qubit neutral-atom quantum computer}, 
      author={Jonathan Wurtz and Alexei Bylinskii and Boris Braverman and Jesse Amato-Grill and Sergio H. Cantu and Florian Huber and Alexander Lukin and Fangli Liu and Phillip Weinberg and John Long and Sheng-Tao Wang and Nathan Gemelke and Alexander Keesling},
      year={2023},
      eprint={2306.11727},
      archivePrefix={arXiv},
      primaryClass={quant-ph}
}

@misc{Hirsbrunner2025,
      title={Quantum criticality and nonequilibrium dynamics on a Lieb lattice of Rydberg atoms}, 
      author={Mark R. Hirsbrunner and Milan Kornjača and Rhine Samajdar and Siva Darbha and Majd Hamdan and Jan Balewski and Ermal Rrapaj and Sheng-Tao Wang and Daan Camps and Fangli Liu and Pedro L. S. Lopes and Katherine Klymko},
      year={2025},
      eprint={2508.05737},
      archivePrefix={arXiv},
      primaryClass={cond-mat.quant-gas},
      url={https://arxiv.org/abs/2508.05737}, 
}

@misc{Darbha2025,
      title={Probing emergent prethermal dynamics and resonant melting on a programmable quantum simulator}, 
      author={Siva Darbha and Alexey Khudorozhkov and Pedro L. S. Lopes and Fangli Liu and Ermal Rrapaj and Jan Balewski and Majd Hamdan and Pavel E. Dolgirev and Alexander Schuckert and Katherine Klymko and Sheng-Tao Wang and Mikhail D. Lukin and Daan Camps and Milan Kornjača},
      year={2025},
      eprint={2510.11706},
      archivePrefix={arXiv},
      primaryClass={quant-ph},
      url={https://arxiv.org/abs/2510.11706}, 
}

@misc{Balewski2025,
      title={Observation of an anomaly in the statistics of Kibble-Zurek defects}, 
      author={Jan Balewski and Alexey Khudorozhkov and Siva Darbha and Omar A. Ashour and Fangli Liu and Ermal Rrapaj and Sheng-Tao Wang and Pedro L. S. Lopes and Katherine Klymko and Milan Kornjača and Daan Camps},
      year={2025},
      eprint={2512.02112},
      archivePrefix={arXiv},
      primaryClass={quant-ph},
      url={https://arxiv.org/abs/2512.02112}, 
}

@Article{Manovitz:2025,
author={Manovitz, Tom and Li, Sophie H. and Ebadi, Sepehr and Samajdar, Rhine and Geim, Alexandra A. and Evered, Simon J. and Bluvstein, Dolev and Zhou, Hengyun and Koyluoglu, Nazli Ugur and Feldmeier, Johannes and Dolgirev, Pavel E. and Maskara, Nishad and Kalinowski, Marcin and Sachdev, Subir and Huse, David A. and Greiner, Markus and Vuleti{\'{c}}, Vladan and Lukin, Mikhail D.},
title={Quantum coarsening and collective dynamics on a programmable simulator},
journal={Nature},
year={2025},
month={Feb},
day={01},
volume={638},
number={8049},
pages={86-92},
issn={1476-4687},
doi={10.1038/s41586-024-08353-5},
url={https://doi.org/10.1038/s41586-024-08353-5}
}

@Article{Cirac2012,
author={Cirac, J. Ignacio
and Zoller, Peter},
title={Goals and opportunities in quantum simulation},
journal={Nature Physics},
year={2012},
month={Apr},
day={01},
volume={8},
number={4},
pages={264-266},
issn={1745-2481},
doi={10.1038/nphys2275},
url={https://doi.org/10.1038/nphys2275}
}

@article{Georgescu2014,
  title = {Quantum simulation},
  author = {Georgescu, I. M. and Ashhab, S. and Nori, Franco},
  journal = {Rev. Mod. Phys.},
  volume = {86},
  issue = {1},
  pages = {153--185},
  numpages = {33},
  year = {2014},
  month = {Mar},
  publisher = {American Physical Society},
  doi = {10.1103/RevModPhys.86.153},
  url = {https://link.aps.org/doi/10.1103/RevModPhys.86.153}
}

@Article{Kornjaca2023,
author={Kornja{\v{c}}a, Milan
and Samajdar, Rhine
and Macr{\`i}, Tommaso
and Gemelke, Nathan
and Wang, Sheng-Tao
and Liu, Fangli},
title={Trimer quantum spin liquid in a honeycomb array of Rydberg atoms},
journal={Communications Physics},
year={2023},
month={Dec},
day={14},
volume={6},
number={1},
pages={358},
issn={2399-3650},
doi={10.1038/s42005-023-01470-z},
url={https://doi.org/10.1038/s42005-023-01470-z}
}

@misc{Mark2025,
      title={Observation of ballistic plasma and memory in high-energy gauge theory dynamics}, 
      author={Daniel K. Mark and Federica M. Surace and Thomas Schuster and Adam L. Shaw and Wenjie Gong and Soonwon Choi and Manuel Endres},
      year={2025},
      eprint={2510.11679},
      archivePrefix={arXiv},
      primaryClass={quant-ph},
      url={https://arxiv.org/abs/2510.11679}, 
}

@article{Samajdar2020,
  title = {Complex Density Wave Orders and Quantum Phase Transitions in a Model of Square-Lattice Rydberg Atom Arrays},
  author = {Samajdar, Rhine and Ho, Wen Wei and Pichler, Hannes and Lukin, Mikhail D. and Sachdev, Subir},
  journal = {Phys. Rev. Lett.},
  volume = {124},
  issue = {10},
  pages = {103601},
  numpages = {7},
  year = {2020},
  month = {Mar},
  publisher = {American Physical Society},
  doi = {10.1103/PhysRevLett.124.103601},
  url = {https://link.aps.org/doi/10.1103/PhysRevLett.124.103601}
}

@Article{Zhang2025,
author={Zhang, Jin
and Cant{\'u}, Sergio H.
and Liu, Fangli
and Bylinskii, Alexei
and Braverman, Boris
and Huber, Florian
and Amato-Grill, Jesse
and Lukin, Alexander
and Gemelke, Nathan
and Keesling, Alexander
and Wang, Sheng-Tao
and Meurice, Yannick
and Tsai, Shan-Wen},
title={Probing quantum floating phases in Rydberg atom arrays},
journal={Nature Communications},
year={2025},
month={Jan},
day={16},
volume={16},
number={1},
pages={712},
issn={2041-1723},
doi={10.1038/s41467-025-55947-2},
url={https://doi.org/10.1038/s41467-025-55947-2}
}

@Article{Chen2023,
author={Chen, Cheng
and Bornet, Guillaume
and Bintz, Marcus
and Emperauger, Gabriel
and Leclerc, Lucas
and Liu, Vincent S.
and Scholl, Pascal
and Barredo, Daniel
and Hauschild, Johannes
and Chatterjee, Shubhayu
and Schuler, Michael
and L{\"a}uchli, Andreas M.
and Zaletel, Michael P.
and Lahaye, Thierry
and Yao, Norman Y.
and Browaeys, Antoine},
title={Continuous symmetry breaking in a two-dimensional Rydberg array},
journal={Nature},
year={2023},
month={Apr},
day={01},
volume={616},
number={7958},
pages={691-695},
issn={1476-4687},
doi={10.1038/s41586-023-05859-2},
url={https://doi.org/10.1038/s41586-023-05859-2}
}

@Article{Scholl2021,
author={Scholl, Pascal
and Schuler, Michael
and Williams, Hannah J.
and Eberharter, Alexander A.
and Barredo, Daniel
and Schymik, Kai-Niklas
and Lienhard, Vincent
and Henry, Louis-Paul
and Lang, Thomas C.
and Lahaye, Thierry
and L{\"a}uchli, Andreas M.
and Browaeys, Antoine},
title={Quantum simulation of 2D antiferromagnets with hundreds of Rydberg atoms},
journal={Nature},
year={2021},
month={Jul},
day={01},
volume={595},
number={7866},
pages={233-238},
issn={1476-4687},
doi={10.1038/s41586-021-03585-1},
url={https://doi.org/10.1038/s41586-021-03585-1}
}

@article{Vidal2003,
  title = {Entanglement in Quantum Critical Phenomena},
  author = {Vidal, G. and Latorre, J. I. and Rico, E. and Kitaev, A.},
  journal = {Phys. Rev. Lett.},
  volume = {90},
  issue = {22},
  pages = {227902},
  year = {2003},
  month = {Jun},
  publisher = {American Physical Society},
  doi = {10.1103/PhysRevLett.90.227902},
  url = {https://link.aps.org/doi/10.1103/PhysRevLett.90.227902}
}

@article{Calabrese2004,
  title = {Entanglement entropy and quantum field theory},
  author = {Calabrese, Pasquale and Cardy, John},
  journal = {Journal of Statistical Mechanics: Theory and Experiment},
  volume = {2004},
  number = {06},
  pages = {P06002},
  year = {2004},
  month = {Jun},
  publisher = {IOP Publishing},
  doi = {10.1088/1742-5468/2004/06/P06002},
  url = {https://doi.org/10.1088/1742-5468/2004/06/P06002}
}

@article{Amico2008,
  title = {Entanglement in many-body systems},
  author = {Amico, Luigi and Fazio, Rosario and Osterloh, Andreas and Vedral, Vlatko},
  journal = {Rev. Mod. Phys.},
  volume = {80},
  issue = {2},
  pages = {517--576},
  year = {2008},
  month = {May},
  publisher = {American Physical Society},
  doi = {10.1103/RevModPhys.80.517},
  url = {https://link.aps.org/doi/10.1103/RevModPhys.80.517}
}

@article{Wierschem2015,
  title = {Origin of modulated phases and magnetic hysteresis in ${\mathrm{TmB}}_{4}$},
  author = {Wierschem, Keola and Sunku, Sai Swaroop and Kong, Tai and Ito, Toshimitsu and Canfield, Paul C. and Panagopoulos, Christos and Sengupta, Pinaki},
  journal = {Phys. Rev. B},
  volume = {92},
  issue = {21},
  pages = {214433},
  numpages = {7},
  year = {2015},
  month = {Dec},
  publisher = {American Physical Society},
  doi = {10.1103/PhysRevB.92.214433},
  url = {https://link.aps.org/doi/10.1103/PhysRevB.92.214433}
}

@article{Ruderman1954,
  title = {Indirect Exchange Coupling of Nuclear Magnetic Moments by Conduction Electrons},
  author = {Ruderman, M. A. and Kittel, C.},
  journal = {Phys. Rev.},
  volume = {96},
  issue = {1},
  pages = {99--102},
  numpages = {0},
  year = {1954},
  month = {Oct},
  publisher = {American Physical Society},
  doi = {10.1103/PhysRev.96.99},
  url = {https://link.aps.org/doi/10.1103/PhysRev.96.99}
}

@article{Huo2013,
    author = {Huo, L. and Huang, W. C. and Yan, Z. B. and Jia, X. T. and Gao, X. S. and Qin, M. H. and Liu, J.-M.},
    title = {The competing spin orders and fractional magnetization plateaus of the classical Heisenberg model on Shastry-Sutherland lattice: Consequence of long-range interactions},
    journal = {Journal of Applied Physics},
    volume = {113},
    number = {7},
    pages = {073908},
    year = {2013},
    month = {02},
    issn = {0021-8979},
    doi = {10.1063/1.4792504},
    url = {https://doi.org/10.1063/1.4792504}
}

@ARTICLE{Villain1980,
  title     = "Order as an effect of disorder",
  author    = "Villain, J and Bidaux, R and Carton, J-P and Conte, R",
  journal   = "Journal Phys.",
  publisher = "EDP Sciences",
  volume    =  41,
  number    =  11,
  pages     = "1263--1272",
  year      =  1980
}

@article{Chubukov1992,
  title = {Order from disorder in a kagom\'e antiferromagnet},
  author = {Chubukov, Andrey},
  journal = {Phys. Rev. Lett.},
  volume = {69},
  issue = {5},
  pages = {832--835},
  numpages = {0},
  year = {1992},
  month = {Aug},
  publisher = {American Physical Society},
  doi = {10.1103/PhysRevLett.69.832},
  url = {https://link.aps.org/doi/10.1103/PhysRevLett.69.832}
}

@article{Chandra1990,
  title = {Ising transition in frustrated Heisenberg models},
  author = {Chandra, P. and Coleman, P. and Larkin, A. I.},
  journal = {Phys. Rev. Lett.},
  volume = {64},
  issue = {1},
  pages = {88--91},
  numpages = {0},
  year = {1990},
  month = {Jan},
  publisher = {American Physical Society},
  doi = {10.1103/PhysRevLett.64.88},
  url = {https://link.aps.org/doi/10.1103/PhysRevLett.64.88}
}

@ARTICLE{Henley1987,
  title     = "Ordering by disorder: Ground-state selection in fcc vector
               antiferromagnets",
  author    = "Henley, Christopher L",
  journal   = "J. Appl. Phys.",
  publisher = "AIP Publishing",
  volume    =  61,
  number    =  8,
  pages     = "3962--3964",
  month     =  apr,
  year      =  1987
  }

@misc{Hudomal2025,
      title={Ergodicity breaking meets criticality in a gauge-theory quantum simulator}, 
      author={Ana Hudomal and Aiden Daniel and Tiago Santiago do Espirito Santo and Milan Kornjača and Tommaso Macrì and Jad C. Halimeh and Guo-Xian Su and Antun Balaž and Zlatko Papić},
      year={2025},
      eprint={2512.23794},
      archivePrefix={arXiv},
      primaryClass={cond-mat.quant-gas},
      url={https://arxiv.org/abs/2512.23794}, 
}

@Article{Fogh2024,
author={Fogh, Ellen
and Nayak, Mithilesh
and Prokhnenko, Oleksandr
and Bartkowiak, Maciej
and Munakata, Koji
and Soh, Jian-Rui
and Turrini, Alexandra A.
and Zayed, Mohamed E.
and Pomjakushina, Ekaterina
and Kageyama, Hiroshi
and Nojiri, Hiroyuki
and Kakurai, Kazuhisa
and Normand, Bruce
and Mila, Fr{\'e}d{\'e}ric
and R{\o}nnow, Henrik M.},
title={Field-induced bound-state condensation and spin-nematic phase in SrCu2(BO3)2 revealed by neutron scattering up to 25.9 T},
journal={Nature Communications},
year={2024},
month={Jan},
day={10},
volume={15},
number={1},
pages={442},
issn={2041-1723},
doi={10.1038/s41467-023-44115-z},
url={https://doi.org/10.1038/s41467-023-44115-z}
}

@Article{deGroot2024,
author={de Groot, Frank M. F.
and Haverkort, Maurits W.
and Elnaggar, Hebatalla
and Juhin, Am{\'e}lie
and Zhou, Ke-Jin
and Glatzel, Pieter},
title={Resonant inelastic X-ray scattering},
journal={Nature Reviews Methods Primers},
year={2024},
month={Jul},
day={04},
volume={4},
number={1},
pages={45},
issn={2662-8449},
doi={10.1038/s43586-024-00322-6},
url={https://doi.org/10.1038/s43586-024-00322-6}
}

@article{Jost2025,
  title = {Time-Resolved X-Ray Spectroscopy from the Atomic Orbital Ground State Up},
  author = {Jost, Daniel and Lomeli, Eder G. and Tang, Ta and Kas, Joshua J. and Rehr, John J. and Lee, Wei-Sheng and Jiang, Hong-Chen and Moritz, Brian and Devereaux, Thomas P.},
  journal = {Phys. Rev. X},
  volume = {15},
  issue = {1},
  pages = {011012},
  numpages = {10},
  year = {2025},
  month = {Jan},
  publisher = {American Physical Society},
  doi = {10.1103/PhysRevX.15.011012},
  url = {https://link.aps.org/doi/10.1103/PhysRevX.15.011012}
}

@article{Mitrano2024,
  title = {Exploring Quantum Materials with Resonant Inelastic X-Ray Scattering},
  author = {Mitrano, M. and Johnston, S. and Kim, Young-June and Dean, M. P. M.},
  journal = {Phys. Rev. X},
  volume = {14},
  issue = {4},
  pages = {040501},
  numpages = {32},
  year = {2024},
  month = {Dec},
  publisher = {American Physical Society},
  doi = {10.1103/PhysRevX.14.040501},
  url = {https://link.aps.org/doi/10.1103/PhysRevX.14.040501}
}

@Article{Bramwell2014,
author={Bramwell, Steven T.
and Keimer, Bernhard},
title={Neutron scattering from quantum condensed matter},
journal={Nature Materials},
year={2014},
month={Aug},
day={01},
volume={13},
number={8},
pages={763-767},
issn={1476-4660},
doi={10.1038/nmat4045},
url={https://doi.org/10.1038/nmat4045}
}

@Article{Daley2022,
author={Daley, Andrew J.
and Bloch, Immanuel
and Kokail, Christian
and Flannigan, Stuart
and Pearson, Natalie
and Troyer, Matthias
and Zoller, Peter},
title={Practical quantum advantage in quantum simulation},
journal={Nature},
year={2022},
month={Jul},
day={01},
volume={607},
number={7920},
pages={667-676},
issn={1476-4687},
doi={10.1038/s41586-022-04940-6},
url={https://doi.org/10.1038/s41586-022-04940-6}
}

@article{Burkard2026,
  title = {Dynamic Correlations of Frustrated Quantum Spins from High-Temperature Expansion},
  author = {Burkard, Ruben and Schneider, Benedikt and Sbierski, Bj\"orn},
  journal = {Phys. Rev. Lett.},
  volume = {136},
  issue = {5},
  pages = {056501},
  numpages = {10},
  year = {2026},
  month = {Feb},
  publisher = {American Physical Society},
  doi = {10.1103/jtjk-x2lw},
  url = {https://link.aps.org/doi/10.1103/jtjk-x2lw}
}

@article{Henelius2000,
  title = {Sign problem in Monte Carlo simulations of frustrated quantum spin systems},
  author = {Henelius, Patrik and Sandvik, Anders W.},
  journal = {Phys. Rev. B},
  volume = {62},
  issue = {2},
  pages = {1102--1113},
  numpages = {0},
  year = {2000},
  month = {Jul},
  publisher = {American Physical Society},
  doi = {10.1103/PhysRevB.62.1102},
  url = {https://link.aps.org/doi/10.1103/PhysRevB.62.1102}
}

@article{Campostrini2001,
  title = {Critical behavior of the three-dimensional $\mathrm{XY}$ universality class},
  author = {Campostrini, Massimo and Hasenbusch, Martin and Pelissetto, Andrea and Rossi, Paolo and Vicari, Ettore},
  journal = {Phys. Rev. B},
  volume = {63},
  issue = {21},
  pages = {214503},
  numpages = {28},
  year = {2001},
  month = {May},
  publisher = {American Physical Society},
  doi = {10.1103/PhysRevB.63.214503},
  url = {https://link.aps.org/doi/10.1103/PhysRevB.63.214503}
}

@article{Lipa2003,
  title = {Specific heat of liquid helium in zero gravity very near the lambda point},
  author = {Lipa, J. A. and Nissen, J. A. and Stricker, D. A. and Swanson, D. R. and Chui, T. C. P.},
  journal = {Phys. Rev. B},
  volume = {68},
  issue = {17},
  pages = {174518},
  numpages = {25},
  year = {2003},
  month = {Nov},
  publisher = {American Physical Society},
  doi = {10.1103/PhysRevB.68.174518},
  url = {https://link.aps.org/doi/10.1103/PhysRevB.68.174518}
}

@article{Yadav2025,
  title   = {Observation of Unprecedented Fractional Magnetization Plateaus in a
             New Shastry-Sutherland Ising Compound},
  author  = {Yadav, Lalit and Rufino, Afonso and Bag, Rabindranath and Ennis, Matthew
             and Koziol, Jan Alexander and dela Cruz, Clarina and Kolesnikov, Alexander I.
             and Garlea, V. Ovidiu and Taddei, Keith M. and Graf, David
             and Schmidt, Kai Phillip and Mila, Fr\'ed\'eric and Haravifard, Sara},
  journal = {Phys. Rev. X}, volume = {15}, pages = {041045}, year = {2025},
  doi = {10.1103/9ynf-xx1t}, eprint = {2405.12405}, archivePrefix = {arXiv}
}

@article{Liu2025,
  title   = {Theory of rare-earth Kramers magnets on a Shastry-Sutherland lattice:
             dimer phases in the presence of strong spin-orbit coupling},
  author  = {Liu, Changle and Duan, Guijing and Yu, Rong},
  journal = {npj Quantum Materials}, volume = {10}, pages = {109}, year = {2025},
  doi = {10.1038/s41535-025-00815-x}
}

@article{Marshall2023,
  title   = {Field-induced partial disorder in a Shastry-Sutherland lattice},
  author  = {Marshall, M. and Billingsley, B. R. and Bai, X. and Ma, Q.
             and Kong, T. and Cao, H.},
  journal = {Nature Communications}, volume = {14}, pages = {3641}, year = {2023},
  doi = {10.1038/s41467-023-39409-1}
}

@Article{Khundzakishvili2025,
AUTHOR = {Khundzakishvili, Guga and Belbase, Bishnu Prasad and Mahendran, Pravin and Zhang, Kevin and Xu, Hanjing and Stoyanoff, Eliana and Checkelsky, Joseph George and Liu, Yaohua and Ye, Linda and Banerjee, Arnab},
TITLE = {Criticality and Magnetic Phases of Ising Shastry–Sutherland Candidate Holmium Tetraboride},
JOURNAL = {Materials},
VOLUME = {18},
YEAR = {2025},
NUMBER = {11},
ARTICLE-NUMBER = {2504},
URL = {https://www.mdpi.com/1996-1944/18/11/2504},
PubMedID = {40508502},
ISSN = {1996-1944},
DOI = {10.3390/ma18112504}
}

\clearpage
\onecolumngrid
\begin{center}
{\bf Supplementary Information: ``Analog quantum simulation of Shastry-Sutherland Ising materials''}
\end{center}

\section{Supplementary theoretical and numerical calculations}
\label{sec:theory_sm}

In this Supplementary Information, we provide the detailed theoretical, numerical, and experimental analysis underlying the main-text results. On the theory side (Sec.~\ref{sec:theory_sm}), we show how the numerical phase diagrams are extracted and validated through entanglement-entropy diagnostics, commensurability analysis, dimer and structure-factor observables for all observed phases, and symmetry-derived Landau theories of the principal phase transitions. On the experimental side (Sec.~\ref{sec:experiment_sm}), we present phase diagrams for all observed phases under different experimental conditions, data that clarify how finite open-array measurements probe bulk ordering tendencies, and the experimental dimer correlations in all observed phases.

\subsection{Entanglement entropy phase diagrams}

In this section, we describe how half-system bipartite entanglement entropy is used to locate phase boundaries and identify strongly quantum regimes in the DMRG phase diagrams. Across a continuous quantum phase transition the correlation length grows, while in gapped phases it is typically smoother and approximately area-law-like. In one-dimensional critical theories one has $S(\ell)\sim \frac{c}{6}\log \xi$ when the correlation length $\xi$ is large but finite, making the growth of entanglement directly tied to the growth of $\xi$~\cite{Vidal2003,Calabrese2004,Amico2008}. Using the entanglement entropy is especially useful in the frustrated Shastry-Sutherland problem, where several nearby ordered states compete and some phase boundaries are hard to identify using local observables. Accordingly, we use ridges in the entanglement-entropy landscape as a first guide to the phase boundaries, and then corroborate the phase assignments with density patterns and structure factors. In addition, regions of large entanglement entropy suggest a quantum nature of the phase, as opposed to simple classically ordered states that are merely dressed by small quantum fluctuations. 

\begin{figure*}[htb]
\centering
\includegraphics[width=0.9\textwidth]{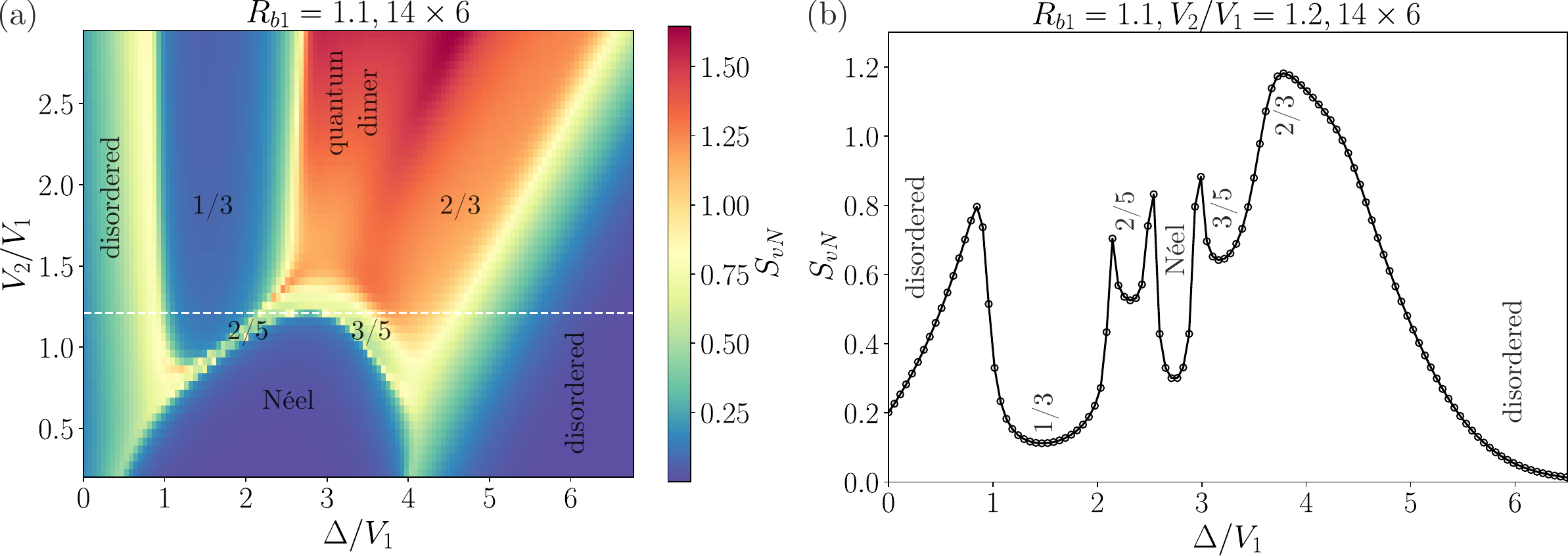}
\caption{\textbf{Entanglement entropy across the Rydberg Shastry-Sutherland lattice phase diagram.} (a) Bipartite von Neumann entanglement entropy, $S_{\mathrm{vN}}$, for the $R_{b1}=1.1$ phase-diagram cut on a $14\times6$ cylinder. Ridges in $S_{\mathrm{vN}}$ identify the phase boundaries, which were used to construct the main-text numerical phase diagram, while the broad high-entanglement region identifies the quantum dimer regime. (b) Entanglement-entropy cut at fixed $V_2/V_1=1.2$, showing the sequence of filling plateaus and peaks associated with the disordered, $1/3$, $2/5$, N\'eel, $3/5$, $2/3$, and high-detuning disordered regimes.
}
\label{fig:FigS1_EE}
\end{figure*}

The entanglement entropy, calculated across a vertical central cut on the Shastry-Sutherland lattice, was mapped across the phase diagram, as shown in Fig.~\ref{fig:FigS1_EE}. The phase boundaries of Fig.~\ref{fig:Fig1_DMRG} are inferred from entanglement entropy landscape. N\'eel, $1/3$ and disordered phases show low entanglement entropy consistent with their classical nature, while $2/5$ ($3/5$) phase shows enhanced entanglement as a signature of quantum fluctuation relevance in the bulk of the phase. The quantum dimer regime exhibits large entanglement entropy, with the chosen cut passing through several dimer bonds. Finally, while the $2/3$ lobe of the $1/3$ ($2/3$) phase is classical, it is incommensurate with the system boundaries, leading to a domain wall in the middle that entangles the left and right domains (see also Fig.~\ref{fig:FigS2_states} for the similar phenomenon in the collinear phase).

\subsection{System size choice and commensurability}

The finite cylinders used in the DMRG calculations must be chosen with care, because several phases in the Shastry-Sutherland phase diagram have large unit cells. A finite open cylinder is commensurate with a candidate ordered phase when the primary ordering wave vector fits the finite geometry without forcing a domain wall. Table~\ref{tab:Tab1S_commensurability} lists the resulting commensurability conditions for the principal ordered phases considered in the main text.

\begin{table}[!htb]
\centering
\begin{tabular}{@{}ccc@{}}
\toprule
phase & commensurate sizes ($N_x$) & main $\mathbf{Q}$-vectors \\ \midrule
N\' eel & $2k$ & $\left(\pi, \pi \right)$ \\
collinear & $2k+1$ & $\left(\pi, 0 \right), \left(0, \pi \right) $ \\
1/3 & $3k-1$ &  $\pm \left(\frac{2\pi}{3}, \pi \right), \pm \left(\pi, \frac{2\pi}{3}\right) $\\
2/3 & $3k+1$ & $\pm \left(\frac{2\pi}{3}, \pi \right), \pm \left(\pi, \frac{2\pi}{3}\right) $ \\
2/5 & $5k-1$ & $\pm \left(\frac{4\pi}{5}, \pi \right), \pm \left(\pi, \frac{4\pi}{5}\right) $ \\
3/5 & $5k+1$ &  $\pm \left(\frac{4\pi}{5}, \pi \right), \pm \left(\pi, \frac{4\pi}{5}\right) $\\ \bottomrule
\end{tabular}
\caption{\textbf{Shastry-Sutherland lattice ordered phases, commensurate system sizes, and ordering wave-vectors.}}
\label{tab:Tab1S_commensurability}
\end{table}
The main numerical phase diagrams reported in Fig.~\ref{fig:Fig1_DMRG} are computed on an ($N_x=14$) cylinder. This length is commensurate with the N\' eel, (1/3), and (2/5) phases, which are the central ordered states discussed in the main text, while keeping the circumference and required DMRG bond dimensions numerically tractable. No single cylinder can be simultaneously commensurate with all phases in the field-induced sequence. In particular, $N_x=14$ is not commensurate with the particle-hole-related ($2/3$) and ($3/5$) plateaus, nor with the collinear state. Those regions can therefore show domain walls or shifted boundaries in finite cylinders. For this reason, phase assignments in the main text are based on the commensurate representatives whenever possible, and are cross-checked using structure factors, entanglement-entropy features, and additional system sizes when commensurability effects are expected to be important.

\subsection{Dimer configurations and structure factors of the phases}

Here, we supplement the phase characterization using real-space densities, dimer-bond diagnostics, and static structure factors. We characterize representative DMRG ground states using both the site-resolved density $\langle n_i\rangle$ and the dimer orientation density on the $V_2$ bonds. For a bond connecting $i$ and its $V_2$ neighbor $\langle i\rangle^2$, the dimer orientation diagnostic is $\langle D_i\rangle=\langle n_i-n_{\langle i\rangle^2}\rangle$. It is large (positive or negative) when the bond contains one Rydberg excitation and one ground-state atom in a definite orientation, and small when the bond is in an equal mixture of these configurations, or the empty/doubly occupied. This makes it a useful complement to the density, as it shows connections between the quantum dimer regime and partially dimerized ordered phases.

\begin{figure*}[htb]
\centering
\includegraphics[width=1.0\textwidth]{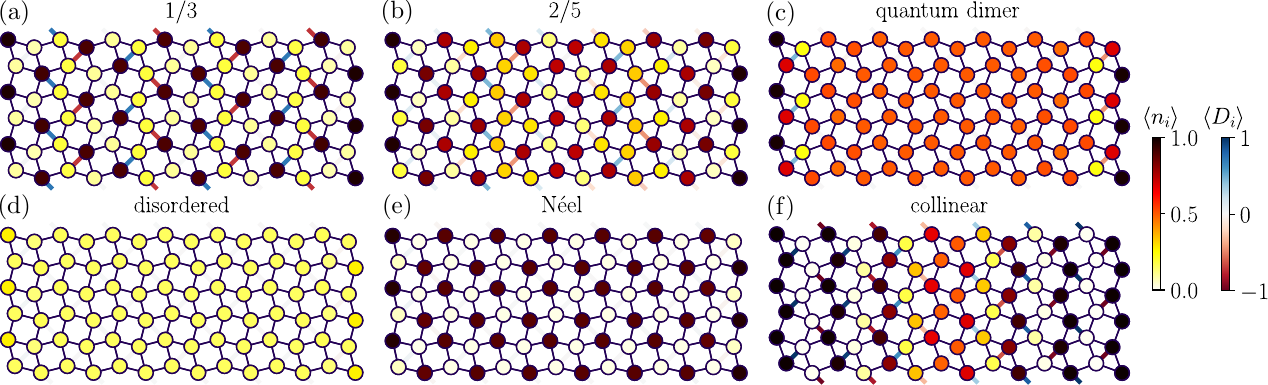}
\caption{\textbf{Density and dimer-density profiles of representative Shastry-Sutherland lattice phases.} Representative DMRG ground states are shown using site-resolved Rydberg densities, $\langle n_i\rangle$, together with the corresponding dimer-density diagnostic $\langle D_i\rangle=\langle n_i-n_{\langle i\rangle^2}\rangle$ on the $V_2$ bonds. Panels show the $1/3$ phase, $2/5$ phase, quantum dimer regime, trivial disordered regime, N\'eel phase, and collinear phase (with a domain wall defect due to commensurability). The ordered phases are visible in the density pattern, and in the ordered dimerization ($1/3$, $2/5$, collinear), while the quantum dimer regime is featureless and most clearly identified by a nearly uniform density together with strong local dimer occupation ($\langle D_i^2 \rangle$, see Fig.~\ref{fig:Fig1_DMRG}).
}
\label{fig:FigS2_states}
\end{figure*}

Figure~\ref{fig:FigS2_states} shows these diagnostics for the main representative phases. The $1/3$ state consists of two N\'eel-like columns followed by one nearly empty column, and the dimer-density pattern reflects that only part of the $V_2$-bond network participates in the ordered configuration. The $2/5$ state has a larger unit cell, with four N\'eel-like columns followed by one nearly empty column, and therefore shows a more weakly modulated but still ordered dimer-density profile. The quantum dimer regime is qualitatively different: its density is nearly uniform and close to $1/2$ in the bulk, the dimer orientation density, $\langle D_i\rangle$  is also featureless and close to zero showing no preference towards dimer orientation, while local dimer occupation, $\langle D_i^2\rangle$ is large on the $V_2$ bonds. The trivial disordered representative is featureless in both diagnostics in the bulk. The N\' eel state is directly visible as a staggered density pattern without any dimerization. In contrast, the collinear state, which is stabilized at higher $R_{b1}$ from the dimer regime, forms columnar antiferromagnetic order with strong stripe contrast in both the density and the dimer-density profile.

\begin{figure*}[htb]
\centering
\includegraphics[width=1.0\textwidth]{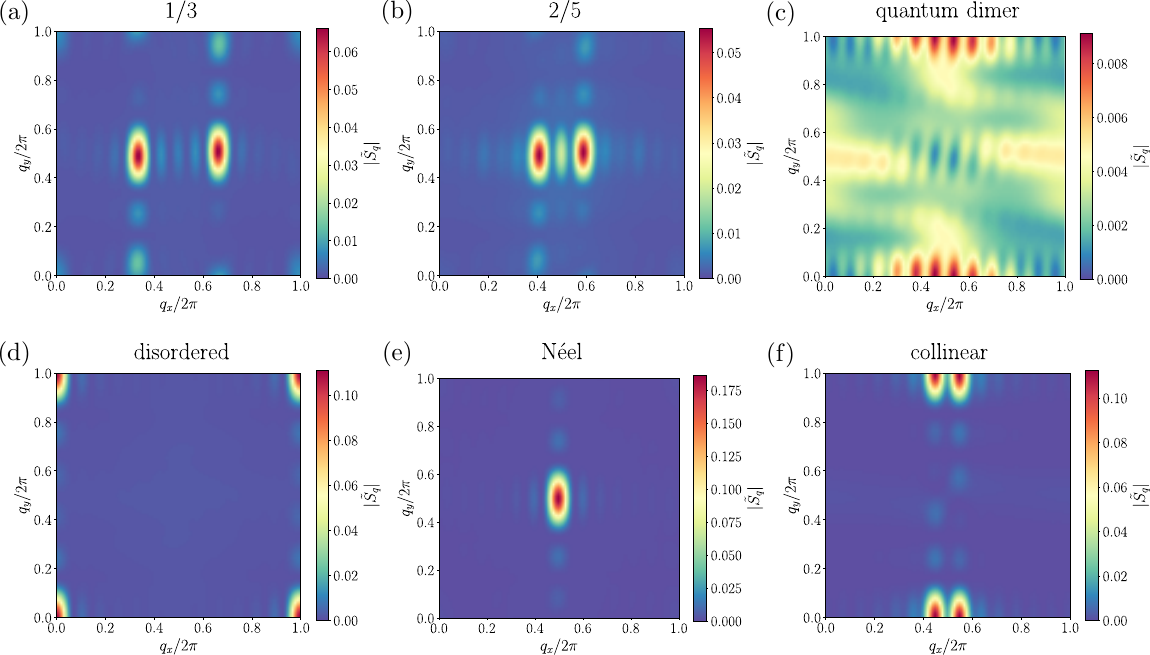}
\caption{\textbf{Numerical structure factors of representative Shastry-Sutherland lattice phases.} Static structure factors are computed from DMRG results using Eq.~\ref{eq:SSF} on the topologically equivalent square lattice. The $1/3$, $2/5$, N\'eel, and collinear phases show peaks at their expected ordering wave vectors (see Tab.~\ref{tab:Tab1S_commensurability}), while the trivial disordered regime lacks a nontrivial sharp peak. The quantum dimer regime instead shows diffuse weight associated with short-range dimer correlations, consistent with the absence of long-range density-wave order. Collinear phase exhibits peak doubling due to the presence of a domain wall defect (see Fig.~\ref{fig:FigS2_states}).
}
\label{fig:FigS2b_SFs}
\end{figure*}

To identify the main ordering wave vectors which serve as the symmetry-sensitive order parameters for the phases, we use the static structure factor defined in Eq.~\ref{eq:SSF}. The positions $\mathbf r'_i$ are taken on the topologically equivalent square lattice, so that the reciprocal-space peaks can be compared directly across all phases. For the ordered states, the expected peaks occur at the primary wave vectors listed in Table~\ref{tab:Tab1S_commensurability}. These peaks are visible in Fig.~\ref{fig:FigS2b_SFs}. By contrast, the disordered state has no nontrivial sharp peak, and the quantum dimer regime shows broad diffuse weight rather than Bragg peaks, consistent with short-range dimer correlations rather than long-range density-wave order. The collinear phase, which is incommensurate with the lattice and exhibits a domain wall in real space, shows a doubling near its expected structure factor peak. In addition to the main (primitive) static structure factor peaks, all ordered phases show secondary peaks that can be formed as multiples and linear combinations of the primitive peaks, but unique characterization of the symmetry-breaking for a given phase can be based on the primitive peaks only.

\subsection{Dimer and flip correlations in the quantum dimer regime}
\label{sec:dimer_flip_correlations}

In this section, we discuss the diagnostics that distinguish the quantum dimer regime from a trivial disordered state or a static dimer crystal. The quantum dimer regime cannot be identified in the same way as the ordered density-wave phases. Its real-space density is nearly featureless in the bulk, and the regular static structure factor does not develop a sharp Bragg peak at a symmetry-breaking ordering vector. Instead, the relevant diagnostics must distinguish three possibilities: a trivial disordered state, a classical dimer or collinear crystal, and a coherent quantum superposition of local dimer configurations.

\begin{figure*}[htb]
\centering
\includegraphics[width=1.0\textwidth]{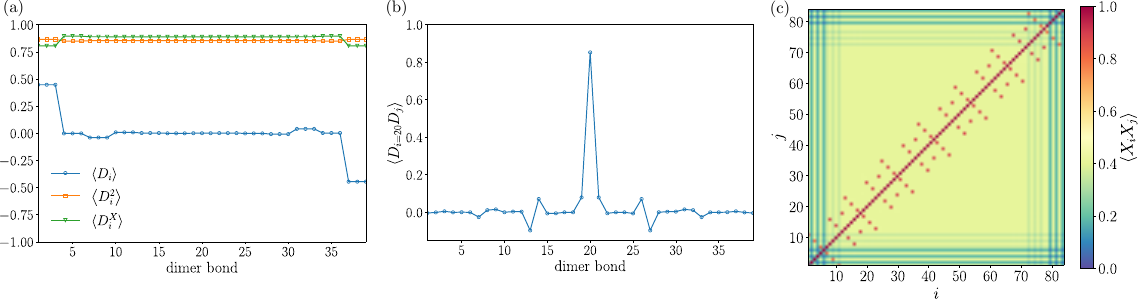}
\caption{\textbf{Numerical dimer and dimer-flip correlations in the quantum dimer regime.}(a) Local dimer diagnostics on $V_2$ bonds. The expectation value $\langle D_i\rangle$ remains small and featureless, while $\langle D_i^2\rangle$ is large in the bulk, showing strong local dimer occupation without selecting a static dimer orientation. The transverse dimer-flip coherence, $\langle D_i^X\rangle=\langle X_iX_{\langle i\rangle^2}\rangle$, is also concentrated on the $V_2$ bonds, providing the direct quantum signature of resonating dimer configurations. (b) Dimer-dimer correlations, $C_{ij}^{D}$, measured relative to a reference dimer. The rapid decay indicates the absence of long-range dimer-crystal order. (c) $\langle X_iX_j \rangle$ coherences, showing pronounced response on the dimer-related pairs, providing a strongly constraining signature of the quantum dimer regime. The transverse coherence is not accessible from native $Z$-basis Rydberg snapshots without an additional basis-rotation or dynamical measurement protocol.
}
\label{fig:FigS3_dimercorrs}
\end{figure*}

We first characterize local dimer occupation on the $V_2$ bonds. For a site $i$ and its $V_2$-bond partner $\langle i\rangle^2$, we define $D_i=n_i-n_{\langle i\rangle^2}$. The two classical dimer configurations on a bond, $\ket{g_i r_{\langle i\rangle^2}}$ and $\ket{r_i g_{\langle i\rangle^2}}$, have $D_i=\pm1$, while the empty and double-excitation configurations give $D_i=0$. As shown in Fig.~\ref{fig:FigS3_dimercorrs}(a), $\langle D_i\rangle$ remains small and featureless in the bulk, indicating that the state does not choose a static dimer orientation. By contrast, $\langle D_i^2\rangle$ is large and nearly uniform on the bulk $V_2$ bonds, showing strong local dimer occupation independent of which of the two classical dimer configurations is present. Local dimer occupation alone does not determine whether the state forms a dimer crystal. We therefore compute the dimer-dimer correlator $C_{ij}^{D}=\langle D_iD_j\rangle$. The correlations in Fig.~\ref{fig:FigS3_dimercorrs}(b) decay rapidly away from a reference dimer, ruling out long-range dimer-crystal order. This distinguishes the quantum dimer regime from the collinear state, where the dimer variables would condense at the corresponding ordering wave vector. The DMRG state is therefore locally dimerized but remains symmetry preserving.

The most direct evidence for the quantum character of this regime comes from transverse dimer-flip coherence. We define the flip operator on a $V_2$ bond as $D_i^X=X_iX_{\langle i\rangle^2}$. This operator connects the two dimer configurations $\ket{g_i r_{\langle i\rangle^2}}$ and $\ket{r_i g_{\langle i\rangle^2}}$. A large value of $\langle D_i^X\rangle$ on the $V_2$ bonds, shown in Fig.~\ref{fig:FigS3_dimercorrs}(a), therefore indicates coherent resonance between the two local dimer states, not just an incoherent mixture of classical dimer coverings. Together with the large $\langle D_i^2\rangle$ and short-ranged $C_{ij}^{D}$, this identifies the regime as a short-range quantum dimer regime rather than a conventional ordered phase. Finally, in  Fig.~\ref{fig:FigS3_dimercorrs}(c), we look at all $X_iX_j$ coherences, showing pronounced response on the dimer-related pairs, $j=\langle i \rangle^2$. This distinguishes the quantum dimer regime from a largely featureless, trivially disordered state and generally provides a strongly constraining signature of it.

These observables also clarify the connection to materials experiments. Diffuse equal-time scattering and bond-sensitive correlations can support the presence of short-range dimerization, while transverse or dynamical probes are needed to establish coherent dimer resonance more conclusively. In transverse-field Shastry-Sutherland materials, such information is in principle accessible through polarized or inelastic neutron scattering, although the required high-field measurements are experimentally demanding~\cite{Flury2025,Fogh2024,Bramwell2014}. The neutral-atom simulator has the complementary limitation that native projective readout is in the $Z$ basis. As a result, density correlations and $\langle D_i^2\rangle$ are directly accessible from snapshots, whereas $\langle X_iX_j\rangle$ requires an additional basis-rotation or dynamical protocol rather than direct measurement, similar in spirit to basis-changing probes used for nonlocal operators in Rydberg-array spin-liquid experiments~\cite{Semeghini2021}.

\subsection{Landau theories of SSL quantum phase transitions}

Here, we derive symmetry-constrained Landau theories for the disorder-order transitions of the principal ordered phases. To characterize the quantum critical points of the Shastry-Sutherland lattice phase diagram, we construct order parameters for the ordered phases and determine how they transform under the lattice symmetries. The resulting symmetry constraints then fix the allowed terms in the Landau theory for the corresponding disorder-order transitions.  The underlying quantum criticalities are non-trivial and could be probed experimentally either through Kibble-Zurek scaling in analog neutral-atom simulators~\cite{Keesling2019, Ebadi2021} or directly in rare-earth tetraborides, providing a particularly sensitive instance of a direct quantum simulation-material connection.

Generally, we write the relevant coarse-grained field as
\begin{equation}\label{eq:phiPsi}
    \phi(\mathbf r)=\mathrm{Re}\left[\sum_n \Psi_n e^{-i\mathbf Q_n\cdot \mathbf r}\right],
\end{equation}
where $\Psi_n$ are the order-parameter amplitudes corresponding to the primary ordering modes. For a lattice symmetry $ T$, the transformation of $\phi(\mathbf r)$ induces a representation $ T_\Psi$ acting in the order-parameter manifold according to:
\begin{equation}\label{eq:symmget}
    T\phi(\mathbf r)
    =
    \mathrm{Re}\left[\sum_n \Psi_n e^{-i\mathbf Q_n\cdot (T\mathbf r)}\right]
    =
    \mathrm{Re}\left[\sum_n ( T_\Psi \Psi_n)e^{-i\mathbf Q_n\cdot \mathbf r}\right].
\end{equation}

\begin{figure*}[htb]
\centering
\includegraphics[width=0.3\textwidth]{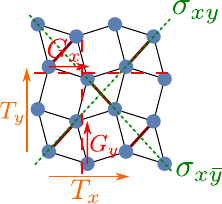}
\caption{\textbf{Symmetry generators for the Shastry-Sutherland lattice. } The nonsymmorphic wallpaper group $p4g$ of the Shastry-Sutherland lattice is generated by two translations ($T$), two reflections ($\sigma$) and two glide planes ($G$), as presented in the figure and described in Eq.~\ref{eq:symmgenerators};
}
\label{fig:FigS4_symmetries}
\end{figure*}
 
To simplify the representation of symmetry generators, we represent the SSL as a topologically equivalent square lattice and use the generators of the nonsymmorphic wallpaper group $p4g$ shown in Fig.~\ref{fig:FigS4_symmetries}. Their action on a square-lattice site $\mathbf r=(x,y)$ is
\begin{align}\label{eq:symmgenerators}
    &T_x:(x,y)\to(x+2,y),\qquad T_y:(x,y)\to(x,y+2),\qquad \sigma_{xy}:(x,y)\to(y,x),\notag\\
    &\sigma_{x\bar y}:(x,y)\to(-y+1,-x+1),\qquad
    G_x:(x,y)\to(x+1,-y),\qquad
    G_y:(x,y)\to(-x,y+1),
\end{align}
where $T$ are translations, $\sigma$ reflections, and $G$ glide planes. For each phase we then construct the invariant polynomial up to quartic order~\cite{Pelissetto2002}. Order--order transitions between the phases found in the SSL phase diagram are generically first order because the competing phases break different symmetries, so our focus is on disorder--order transitions.

\subsubsection*{Disorder--N\'eel}

The N\'eel phase is described by a single ordering wave-vector and a single real order parameter,
\begin{equation}\label{eq:phiNeel}
    \phi(\mathbf r)=\mathrm{Re}\left[\Psi e^{-i(\pi,\pi)\cdot \mathbf r}\right].
\end{equation}
Applying Eqs.~\ref{eq:symmget} and \ref{eq:symmgenerators} gives the standard Ising theory,
\begin{equation}\label{eq:LNeel}
    \bar{\mathcal L}=r\Psi^2+u\Psi^4.
\end{equation}
Thus, the disorder--N\'eel transition is expected to lie in the $(2+1)$D Ising universality class~\cite{Pelissetto2002, El-Showk2014}, as explored extensively in the square-lattice case with analog neutral atom simulators~\cite{Ebadi2021}.

\subsubsection*{Disorder--1/3}

The $1/3$ ($2/3$) phase is described by two complex order parameters,
\begin{equation}\label{eq:phi13}
    \phi(\mathbf r)=
    \mathrm{Re}\left[
        \Psi_1 e^{-i(2\pi/3,\pi)\cdot \mathbf r}
        +
        \Psi_2 e^{-i(\pi,2\pi/3)\cdot \mathbf r}
    \right].
\end{equation}
Since this is the first nontrivial example, we give the symmetry action explicitly. Taking $\xi=e^{2\pi i/3}$, one finds
\begin{align}\label{eq:mat13compact}
    T_x:\;(\Psi_1,\Psi_2)&\to(\xi \Psi_1,\Psi_2), \qquad
    &T_y:\;(\Psi_1,\Psi_2)\to(\Psi_1,\xi \Psi_2), \qquad
    \sigma_{xy}:\;(\Psi_1,\Psi_2)&\to(\Psi_2,\Psi_1),\notag\\
    \sigma_{x\bar y}:\;(\Psi_1,\Psi_2)&\to(-\xi \Psi_2^*,-\xi \Psi_1^*), \qquad
    &G_x:\;(\Psi_1,\Psi_2)\to(\xi^*\Psi_1,-\Psi_2^*), \qquad
    G_y:\;(\Psi_1,\Psi_2)&\to(-\Psi_1^*,\xi^*\Psi_2).
\end{align}
The corresponding invariant polynomial is
\begin{equation}\label{eq:L13}
    \bar{\mathcal L}
    =
    r\left(|\Psi_1|^2+|\Psi_2|^2\right)
    +
    u\left(|\Psi_1|^4+|\Psi_2|^4\right)
    +
    v|\Psi_1|^2|\Psi_2|^2
    +
    \lambda\,\mathrm{Im}\left(\Psi_1^3+\Psi_2^3\right).
\end{equation}
The cubic invariant is the defining feature of this transition and identifies the theory as two coupled three-state Potts fields with quartic coupling. Since even the single-field $(2+1)$D Potts transition is weakly first-order~\cite {Janke1997}, this coupled theory could constitute an interesting experimental target. Thus, while the $1/3$ phase is a well-known classically stabilized SSL phase, the effect of strong quantum fluctuations at the disordered-$1/3$ phase transition could provide insights into non-trivial quantum critical phenomena.

\subsubsection*{Disorder--collinear}

The collinear phase can be viewed as the condensation channel out of the symmetry-preserving quantum dimer regime and is described by two real order parameters,
\begin{equation}\label{eq:phicoll}
    \phi(\mathbf r)=
    \mathrm{Re}\left[
        \Psi_1 e^{-i(\pi,0)\cdot \mathbf r}
        +
        \Psi_2 e^{-i(0,\pi)\cdot \mathbf r}
    \right].
\end{equation}
The symmetry analysis gives the quartic Landau polynomial
\begin{equation}\label{eq:Lcoll}
    \bar{\mathcal L}
    =
    r(\Psi_1^2+\Psi_2^2)
    +
    u(\Psi_1^2+\Psi_2^2)^2
    +
    v(\Psi_1^4+\Psi_2^4).
\end{equation}
Requiring condensation of only one of the two components gives the conditions $u+v>0$ and $v<0$. This is the $D_4$ anisotropic $O(2)$ theory~\cite{Aharony1973}. For $v<0$ and $v>-2u/3$, the flow runs to the stable XY fixed point~\cite{Korzhenevskii1975, Pelissetto2002}, corresponding to a second-order disorder--collinear transition in the $(2+1)$D XY universality class. This universality class
also governs the finite-temperature superfluid transition in liquid He-4, raising a possible connection between SSL physics and one of the fundamental quantum many-body phenomena~\cite{Campostrini2001, Lipa2003}. For $-u<v<-2u/3$, the flow is driven away from criticality, indicating a fluctuation-induced first-order transition. This latter regime is likely the one realized in the corresponding disorder--striated transition of the square-lattice Rydberg model~\cite{Kalinowski2022}.

\subsubsection*{Disorder--2/5}

The $2/5$ ($3/5$) phase is described by two complex order parameters,
\begin{equation}\label{eq:phi25}
    \phi(\mathbf r)=
    \mathrm{Re}\left[
        \Psi_1 e^{-i(4\pi/5,\pi)\cdot \mathbf r}
        +
        \Psi_2 e^{-i(\pi,4\pi/5)\cdot \mathbf r}
    \right].
\end{equation}
By the same symmetry-analysis procedure, one obtains the quartic invariant polynomial
\begin{equation}\label{eq:L25}
    \bar{\mathcal L}
    =
    r\left(|\Psi_1|^2+|\Psi_2|^2\right)
    +
    u\left(|\Psi_1|^4+|\Psi_2|^4\right)
    +
    v|\Psi_1|^2|\Psi_2|^2.
\end{equation}
Unlike the $1/3$ case, the first symmetry-allowed phase-locking term appears only at fifth order. The ordered $2/5$ state again requires $v>0$ and $v>2u$ (or equivalently, $u>0$ and $v>-2u$). The resulting theory is the $MN$ model with $M=N=2$~\cite{Grinstein1976}, equivalently the $O(2)\times O(2)$ chiral model~\cite{Kawamura1988}, analogous to the disorder--banded transition in the square-lattice Rydberg model~\cite{Samajdar2020}. More generally, the same quartic Landau theory describes all disorder--$k/(2k+1)$ transitions with $k\ge 2$, with differences between the plateaus appearing only in higher-order, irrelevant invariants.

\section{Supplementary experimental results}
\label{sec:experiment_sm}
\subsection{Phase diagrams of different order parameters under varying boundary pinning}

\begin{figure*}[!htb]
\centering
\includegraphics[width=\textwidth]{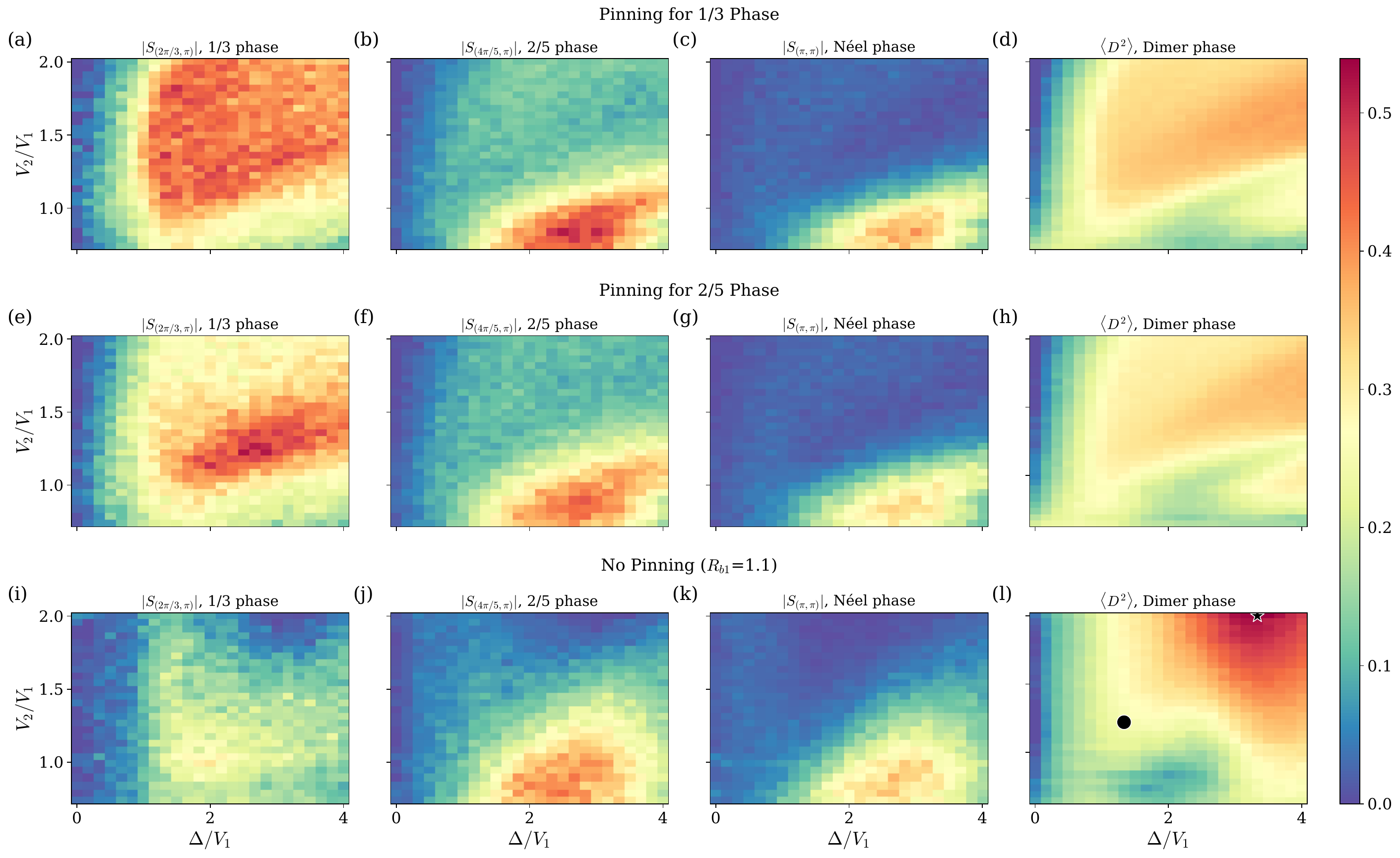}
\caption{\textbf{Experimental phase diagrams under varying boundary-pinning conditions.}
Experimentally extracted order-parameter maps at fixed $R_{b1}=1.1$ for three boundary-pinning configurations. (a)-(d) correspond to $1/3$-favoring pinning, (e)-(h) $2/5$-favoring pinning, and the unpinned geometry is shown in (i)-(l). Columns show, respectively, the $1/3$ structure-factor response, the $2/5$ structure-factor response, the N\'eel structure-factor response, and the average local dimer occupation $\langle D^2\rangle$. The same qualitative organization of the phase diagram is visible across boundary choices, while the contrast and apparent extent of each phase depend on whether the boundary condition favors the corresponding order. This shows that boundary pinning biases the finite open array toward a given ordering pattern and enhances the contrast of its corresponding diagnostic. The markers in panel (l) correspond to parameter points used in Fig.~\ref{fig:Fig_exp_dimercorrs}.}
\label{fig:Fig_suppl}
\end{figure*}

In the main text, we present the experimental response maps using the boundary condition that most clearly reveals each phase. Here, we present a complementary analysis in which the four-phase order-parameter diagnostics are evaluated for each of the three boundary-pinning configurations used experimentally. Figure~\ref{fig:Fig_suppl} is organized by pinning condition: the first row uses $1/3$-favoring boundary atoms, the second row uses $2/5$-favoring boundary atoms, and the third row corresponds to the native unpinned geometry, which favors N\'eel order through the natural reduced-coordination structure of the open boundary. Within each row, we plot the $1/3$ structure factor, the $2/5$ structure factor, the N\'eel structure factor, and the average local dimer occupation $\langle D^2\rangle$. This data separates two effects that are combined in the main-text phase diagrams: the intrinsic location of a phase in the $(\Delta/V_1,V_2/V_1)$ plane and the degree to which a particular open boundary enhances its visibility in a finite system.

The main result is that the same qualitative phase organization persists across all pinning choices. The $1/3$ response is largest under $1/3$-adapted pinning, but a broad $1/3$ region remains visible also under the $2/5$ and unpinned boundary conditions. Similarly, the $2/5$ response is enhanced by $2/5$-adapted pinning, but it is not confined to that boundary choice: a finite response appears in the same central frustrated region near the N\'eel--$1/3$ boundary where the numerical phase diagram places the fluctuation-stabilized $2/5$ phase. The N\'eel order parameter is strongest in the unpinned geometry, consistent with the natural tendency of open edges to favor staggered occupation, but the N\'eel region remains localized to the low-$V_2/V_1$ part of the phase diagram across pinning choices. Finally, the dimer diagnostic is comparatively insensitive to the boundary condition, reflecting the fact that the quantum dimer regime is characterized by local bond correlations rather than a long-range symmetry-breaking pattern consistent with the boundary.

This cross-pinning comparison is important for interpreting finite-array data. If the auxiliary atoms simply imprinted a target pattern, one would expect a given phase diagnostic to appear only, or overwhelmingly, under its matching boundary condition. Instead, the data show a more constrained behavior: boundary pinning changes the contrast and apparent extent of a phase, but does not move its response to arbitrary regions of parameter space. The largest changes occur near phase boundaries and in the larger unit cell orders ($1/3,\, 2/5$). The region of enhanced $2/5$ response also shows considerable overlap with N\'eel region, due to ordering wave vectors that are close within the finite experimental momentum resolution. In contrast, the dimer regime remains robust because its diagnostic is local and does not rely on resolving a sharp structure factor peak. Thus, the boundary atoms act as symmetry-selective finite-size biases: they modify the contrast and apparent extent of a response, while the location of that response remains strongly constrained by the underlying Hamiltonian parameters. This cross-pinning comparison does not by itself establish the corresponding thermodynamic phases, but it provides a useful check that the observed patterns are not determined solely by the chosen boundary configuration.

\subsection{Experimental dimer correlations}

In this subsection, we present experimentally accessible dimer-correlation diagnostics for the quantum dimer regime and compare them to the $1/3$ density-wave phase.

\begin{figure*}[htb]
\centering
\includegraphics[width=\textwidth]{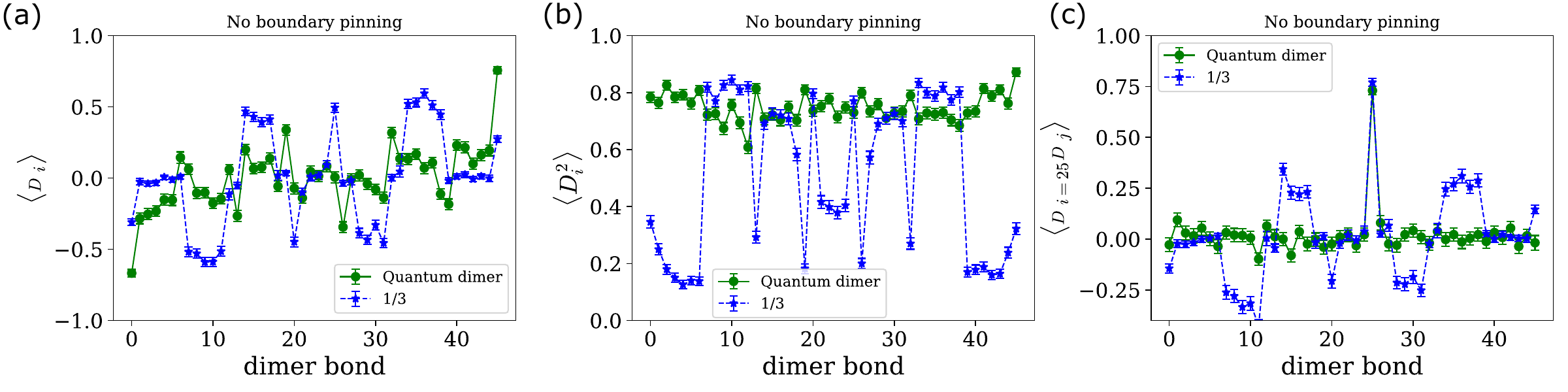}
\caption{\textbf{Experimental dimer correlations.}
Experimentally measured $Z$-basis dimer diagnostics for a representative quantum dimer point (green, $V_2/V_1=2.0$, $\Delta/V_1=3.33$) and a representative $1/3$ point (blue, $V_2/V_1=1.725$, $\Delta/V_1=1.5$), both measured without additional boundary pinning (see Fig.~\ref{fig:Fig_suppl}). (a) Local dimer orientation $\langle D_i\rangle$ on the $V_2$ bonds. (b) Local dimer occupation $\langle D_i^2\rangle$. (c) Dimer-dimer correlator $\langle D_{i=25}D_j\rangle$ measured relative to the reference dimer bond $i=25$. The quantum dimer regime shows weak static dimer orientation together with strong, comparatively uniform local dimer occupation, whereas the $1/3$ phase shows bond-dependent modulation inherited from its density order.}
\label{fig:Fig_exp_dimercorrs}
\end{figure*}

The experimental snapshots allow us to test the local dimer diagnostics introduced in Sec.~\ref{sec:dimer_flip_correlations}, but only in the Rydberg-density basis. Figure~\ref{fig:Fig_exp_dimercorrs} extends the main-text identification of the quantum dimer regime in Figs.~\ref{fig:Fig3_phases}(d,h) and~\ref{fig:Fig4_PDs}(d), where the density is nearly featureless, the static structure factor lacks a sharp ordering peak, and the phase diagram is instead characterized by the averaged dimer occupation $\langle D^2\rangle$. Here, we compare the underlying bond-resolved diagnostics at representative points in the quantum dimer regime (green) and the $1/3$ phase (blue), both in the unpinned geometry. In Fig.~\ref{fig:Fig_exp_dimercorrs}(a), the local dimer orientation $\langle D_i\rangle$ remains comparatively featureless in the dimer regime, consistent with the absence of classical dimer order. The residual profile is a finite-open-boundary effect: $\langle D_i\rangle$ is close to zero in the bulk, but the open edges induce a finite local orientation whose influence weakly penetrates inward. Because the resulting deviation from zero is small and changes sign approximately antisymmetrically about the center, it appears roughly linear when plotted against the dimer-bond index. In Fig.~\ref{fig:Fig_exp_dimercorrs}(b), the corresponding occupation diagnostic $\langle D_i^2\rangle$ is large and nearly uniform across the bulk $V_2$ bonds, showing strong local dimer occupation. By contrast, the $1/3$ data show stronger bond-to-bond modulation, reflecting the underlying density wave rather than a uniform dimerized background.

The dimer-dimer correlator in Fig.~\ref{fig:Fig_exp_dimercorrs}(c), measured relative to a representative reference bond, provides a complementary comparison between the two regimes. The quantum dimer data show a fast decay across the dimer bonds, consistent with the diffuse structure-factor weight shown in Fig.~\ref{fig:Fig3_phases}(h), while the $1/3$ phase inherits the spatial modulation of the ordered density pattern. These measurements therefore support the interpretation used in the main text: the experimental dimer regime is not identified by a sharp density structure-factor peak, but by strong local $V_2$-bond occupation together with the absence of static dimer-orientation order. The remaining, intrinsically quantum diagnostic, transverse dimer-flip coherence, is not directly accessible from the native density snapshots and would require an additional basis-rotation or dynamical measurement protocol~\cite{Semeghini2021, geim2026engineeringquantumcriticalitydynamics}.

\end{document}